\pdfoutput=1

\documentclass[12pt,a4paper]{article}

\usepackage{ifthen} 
\usepackage{rotating}
\usepackage{dashrule}
\usepackage{array} 
\usepackage{dcolumn}
\usepackage{comment}
\usepackage{xcolor}

\usepackage{dashrule} 
\usepackage{graphpap} 
\usepackage{rotating} 
\usepackage{tikz}
\usepackage{xcolor}
\usepackage{longtable} 

\newboolean{pdflatex}
\setboolean{pdflatex}{true} 

\newboolean{articletitles}
\setboolean{articletitles}{true} 

\newboolean{uprightparticles}
\setboolean{uprightparticles}{True} 

\def\paperauthors{LHCb collaboration} 
\def\paperasciititle{Observation of the Lambda_b -> chi_c1(3872) proton K- decay} 
\def\papertitle{Observation of the \LbToXPK decay} 
\def\paperkeywords{{High Energy Physics}, {LHCb}} 
\def\papercopyright{\the\year\ CERN for the benefit of the LHCb collaboration} 
\def\paperlicence{CC-BY-4.0 licence}
\def\paperlicenceurl{https://creativecommons.org/licenses/by/4.0/}

\usepackage[top=1in, bottom=1.25in, left=1in, right=1in]{geometry}

\usepackage{microtype}
\usepackage{lineno}  
\usepackage{xspace} 
\usepackage{caption} 

\usepackage{graphicx}  
\usepackage{color}
\usepackage{colortbl}
\graphicspath{{./figs/}} 
\DeclareGraphicsExtensions{.pdf,.PDF,png,.PNG}

\usepackage{amsmath} 
\usepackage{amssymb}
\usepackage{amsfonts}
\usepackage{upgreek} 

\newcommand*\patchAmsMathEnvironmentForLineno[1]{%
\expandafter\let\csname old#1\expandafter\endcsname\csname #1\endcsname
\expandafter\let\csname oldend#1\expandafter\endcsname\csname
end#1\endcsname
 \renewenvironment{#1}%
   {\linenomath\csname old#1\endcsname}%
   {\csname oldend#1\endcsname\endlinenomath}%
}
\newcommand*\patchBothAmsMathEnvironmentsForLineno[1]{%
  \patchAmsMathEnvironmentForLineno{#1}%
  \patchAmsMathEnvironmentForLineno{#1*}%
}
\AtBeginDocument{%
\patchBothAmsMathEnvironmentsForLineno{equation}%
\patchBothAmsMathEnvironmentsForLineno{align}%
\patchBothAmsMathEnvironmentsForLineno{flalign}%
\patchBothAmsMathEnvironmentsForLineno{alignat}%
\patchBothAmsMathEnvironmentsForLineno{gather}%
\patchBothAmsMathEnvironmentsForLineno{multline}%
\patchBothAmsMathEnvironmentsForLineno{eqnarray}%
}

\usepackage{hyperxmp}

\usepackage[pdftex,
            pdfauthor={\paperauthors},
            pdftitle={\paperasciititle},
            pdfkeywords={\paperkeywords},
            pdfcopyright={Copyright (C) \papercopyright},
            pdflicenseurl={\paperlicenceurl}]{hyperref}

\usepackage[colorinlistoftodos,textsize=scriptsize]{todonotes}

\usepackage[all]{hypcap} 

\usepackage{xspace} 
\usepackage{upgreek}

\def\lhcb   {\mbox{LHCb}\xspace}

\def\belle  {\mbox{Belle}\xspace}

\def\MagUp {\mbox{\em Mag\kern -0.05em Up}\xspace}

\ifthenelse{\boolean{uprightparticles}}%
{

 \def\Pmu         {\ensuremath{\upmu}\xspace}

 \def\Ppi         {\ensuremath{\uppi}\xspace}                 
                  
 \def\Prho        {\ensuremath{\uprho}\xspace}

 \def\Pphi        {\ensuremath{\upphi}\xspace}                 
                  
 \def\Pchi        {\ensuremath{\upchi}\xspace}                 
 \def\Ppsi        {\ensuremath{\uppsi}\xspace}

 \def\PDelta      {\ensuremath{\Delta}\xspace}                 
 \def\PXi         {\ensuremath{\Xi}\xspace}                 
 \def\PLambda     {\ensuremath{\Lambda}\xspace}                 
 \def\PSigma      {\ensuremath{\Sigma}\xspace}                 
 \def\POmega      {\ensuremath{\Omega}\xspace}                 
 \def\PUpsilon    {\ensuremath{\Upsilon}\xspace}

 \def\PB      {\ensuremath{\mathrm{B}}\xspace}                 
                  
 \def\PD      {\ensuremath{\mathrm{D}}\xspace}

 \def\PJ      {\ensuremath{\mathrm{J}}\xspace}                 
 \def\PK      {\ensuremath{\mathrm{K}}\xspace}

 \def\PX      {\ensuremath{\mathrm{X}}\xspace}

 \def\Pb      {\ensuremath{\mathrm{b}}\xspace}                 
 \def\Pc      {\ensuremath{\mathrm{c}}\xspace}

 \def\Pi      {\ensuremath{\mathrm{i}}\xspace}

 \def\Pp      {\ensuremath{\mathrm{p}}\xspace}

 \def\Ps      {\ensuremath{\mathrm{s}}\xspace}

 \def\thebaroffset{0.0em}
}
{

 \def\Pmu         {\ensuremath{\mu}\xspace}

 \def\Ppi         {\ensuremath{\pi}\xspace}                 
                  
 \def\Prho        {\ensuremath{\rho}\xspace}

 \def\Pphi        {\ensuremath{\phi}\xspace}                 
                  
 \def\Pchi        {\ensuremath{\chi}\xspace}                 
 \def\Ppsi        {\ensuremath{\psi}\xspace}                 
                  
 \mathchardef\PDelta="7101
 \mathchardef\PXi="7104
 \mathchardef\PLambda="7103
 \mathchardef\PSigma="7106
 \mathchardef\POmega="710A
 \mathchardef\PUpsilon="7107
                  
 \def\PB      {\ensuremath{B}\xspace}                 
                  
 \def\PD      {\ensuremath{D}\xspace}

 \def\PJ      {\ensuremath{J}\xspace}                 
 \def\PK      {\ensuremath{K}\xspace}

 \def\PX      {\ensuremath{X}\xspace}

 \def\Pb      {\ensuremath{b}\xspace}                 
 \def\Pc      {\ensuremath{c}\xspace}

 \def\Pi      {\ensuremath{i}\xspace}

 \def\Pp      {\ensuremath{p}\xspace}

 \def\Ps      {\ensuremath{s}\xspace}

 \def\thebaroffset{0.18em}
}
\newcommand{\offsetoverline}[2][\thebaroffset]{\kern #1\overline{\kern -#1 #2}}%

\makeatletter
\ifcase \@ptsize \relax
  \newcommand{\miniscule}{\@setfontsize\miniscule{4}{5}}
\or
  \newcommand{\miniscule}{\@setfontsize\miniscule{5}{6}}
\or
  \newcommand{\miniscule}{\@setfontsize\miniscule{5}{6}}
\fi
\makeatother

\DeclareRobustCommand{\optbar}[1]{\shortstack{{\miniscule (\rule[.5ex]{1.25em}{.18mm})}
  \\ [-.7ex] $#1$}}

\def\mumu       {{\ensuremath{\Pmu^+\Pmu^-}}\xspace}

\def\squark    {{\ensuremath{\Ps}}\xspace}

\def\cquark    {{\ensuremath{\Pc}}\xspace}
\def\cquarkbar {{\ensuremath{\overline \cquark}}\xspace}
\def\ccbar     {{\ensuremath{\cquark\cquarkbar}}\xspace}
\def\bquark    {{\ensuremath{\Pb}}\xspace}

\def\pion   {{\ensuremath{\Ppi}}\xspace}

\def\pip    {{\ensuremath{\pion^+}}\xspace}
\def\pim    {{\ensuremath{\pion^-}}\xspace}

\def\kaon    {{\ensuremath{\PK}}\xspace}

\def\KorKbar {\kern \thebaroffset\optbar{\kern -\thebaroffset \PK}{}\xspace}

\def\Kp      {{\ensuremath{\kaon^+}}\xspace}
\def\Km      {{\ensuremath{\kaon^-}}}

\def\KS      {{\ensuremath{\kaon^0_{\mathrm{S}}}}\xspace}

\def\D       {{\ensuremath{\PD}}\xspace}

\def\DorDbar {\kern \thebaroffset\optbar{\kern -\thebaroffset \PD}\xspace}
\def\Dz      {{\ensuremath{\D^0}}\xspace}

\def\Dstarp  {{\ensuremath{\D^{*+}}}\xspace}

\def\Ds      {{\ensuremath{\D^+_\squark}}\xspace}

\def\B       {{\ensuremath{\PB}}\xspace}

\def\BorBbar {\kern \thebaroffset\optbar{\kern -\thebaroffset \PB}\xspace}
\def\Bz      {{\ensuremath{\B^0}}\xspace}

\def\Bd      {{\ensuremath{\B^0}}\xspace}

\def\BdorBdbar {\kern \thebaroffset\optbar{\kern -\thebaroffset \Bd}\xspace}

\def\Bs      {{\ensuremath{\B^0_\squark}}\xspace}

\def\BsorBsbar {\kern \thebaroffset\optbar{\kern -\thebaroffset \Bs}\xspace}

\def\x  {{\ensuremath{\PX{(3872)}}}\xspace}

\def\jpsi     {{\ensuremath{{\PJ\mskip -3mu/\mskip -2mu\Ppsi\mskip 2mu}}}\xspace}
\def\psitwos  {{\ensuremath{\Ppsi{(\rm{2S})}}}\xspace}

\def\chiconex  {{\ensuremath{\Pchi_{\cquark 1}(3872)}}\xspace}

\def\Y#1S{\ensuremath{\PUpsilon{(#1S)}}\xspace}

\def\proton      {{\ensuremath{\Pp}}\xspace}

\def\Lz          {{\ensuremath{\PLambda}}\xspace}

\def\LorLbar     {\kern \thebaroffset\optbar{\kern -\thebaroffset \PLambda}\xspace}
\def\Lambdares   {{\ensuremath{\PLambda(1520)}}\xspace}

\def\Lc          {{\ensuremath{\Lz^+_\cquark}}\xspace}

\def\Lb           {{\ensuremath{\Lz^0_\bquark}}\xspace}

\def\BF         {{\ensuremath{\mathcal{B}}}\xspace}
\def\BR         {\BF}

\newcommand{\decay}[2]{\ensuremath{#1\!\to #2}\xspace}         

\def\to                 {\ensuremath{\rightarrow}\xspace}

\def\LbToLamst    {\decay{\Lb}{\chiconex\Lambdares}}

\def\LbToXPK      {\decay{\Lb}{\chiconex\proton\Km}}

\def\LbToNR       {\decay{\Lb}{\jpsi\pip\pim\proton\Km}}
\def\LbToJpsiPK   {\decay{\Lb}{\jpsi\proton\Km}}
\def\LbToPsiPK    {\decay{\Lb}{\psitwos\proton\Km}}
\def\XToJPsipipi  {\decay{\chiconex}{\jpsi\pip\pim}}
\def\PsiToJPsipipi{\decay{\psitwos}{\jpsi\pip\pim}}

\def\JpsiPiPi     {{\jpsi\pip\pim}~}
\def\JpsiMuMu    {\decay{\jpsi}{\mumu}}

\def\BzTojpsipipipK   {{\decay{\Bz}{\jpsi\pip\pim\pip\Km}\xspace}}

\def\DstarPlus   {{\decay{\Dstarp}{\Dz(\to\Km\pip)\pip}}}
\def\Dsplus      {{\decay{\Ds}{\Pphi(\to\Kp\Km)\pip}}}
\def\Ksz         {{\decay{\KS}{\pip\pim}}}

\def\Lcplus      {{\decay{\Lc}{\proton\Kp\pim}}}

\def\JpsiPiPiPK {{\jpsi\pip\pim\proton\Km}~}

\def\AT#1     {\ensuremath{A_{\mathrm{T}}^{#1}}\xspace}           

\def\C#1      {\ensuremath{\mathcal{C}_{#1}}\xspace}                       
\def\Cp#1     {\ensuremath{\mathcal{C}_{#1}^{'}}\xspace}                    
\def\Ceff#1   {\ensuremath{\mathcal{C}_{#1}^{\mathrm{(eff)}}}\xspace}        
\def\Cpeff#1  {\ensuremath{\mathcal{C}_{#1}^{'\mathrm{(eff)}}}\xspace}       
\def\Ope#1    {\ensuremath{\mathcal{O}_{#1}}\xspace}                       
\def\Opep#1   {\ensuremath{\mathcal{O}_{#1}^{'}}\xspace}                    

\newcommand{\nospaceunit}[1]{\ensuremath{\text{#1}}}       
\newcommand{\aunit}[1]{\ensuremath{\text{\,#1}}}       

\newcommand{\tev}{\aunit{Te\kern -0.1em V}\xspace}
\newcommand{\gev}{\aunit{Ge\kern -0.1em V}\xspace}
\newcommand{\mev}{\aunit{Me\kern -0.1em V}\xspace}
\newcommand{\kev}{\aunit{ke\kern -0.1em V}\xspace}
\newcommand{\ev}{\aunit{e\kern -0.1em V}\xspace}
\newcommand{\mevc}{\ensuremath{\aunit{Me\kern -0.1em V\!/}c}\xspace}
\newcommand{\gevc}{\ensuremath{\aunit{Ge\kern -0.1em V\!/}c}\xspace}
\newcommand{\mevcc}{\ensuremath{\aunit{Me\kern -0.1em V\!/}c^2}\xspace}
\newcommand{\gevcc}{\ensuremath{\aunit{Ge\kern -0.1em V\!/}c^2}\xspace}

\def\mum  {\ensuremath{\,\upmu\nospaceunit{m}}\xspace}

\def\fb   {\ensuremath{\aunit{fb}}\xspace}
\def\invfb   {\ensuremath{\fb^{-1}}\xspace}

\newcommand{\chisq}{\ensuremath{\chi^2}\xspace}

\newcommand{\chisqip}{\ensuremath{\chi^2_{\text{IP}}}\xspace}

\def\gsim{{~\raise.15em\hbox{$>$}\kern-.85em
          \lower.35em\hbox{$\sim$}~}\xspace}
\def\lsim{{~\raise.15em\hbox{$<$}\kern-.85em
          \lower.35em\hbox{$\sim$}~}\xspace}

\def\pt         {\ensuremath{p_{\mathrm{T}}}\xspace}

\def\evtgen     {\mbox{\textsc{EvtGen}}\xspace}

\def\geant      {\mbox{\textsc{Geant4}}\xspace}

\def\photos     {\mbox{\textsc{Photos}}\xspace}

\def\pythia     {\mbox{\textsc{Pythia}}\xspace}

\xspace

\def\tell1  {TELL1\xspace}
\def\ukl1   {UKL1\xspace}

\newcommand{\etc}{\mbox{\itshape etc.}\xspace}

\usepackage{cite} 
\usepackage{mciteplus}

\usepackage{longtable} 

\newcolumntype{d}[1]{D{,}{\,\pm\,}{#1} }
\newcolumntype{f}[1]{D{,}{.}{#1} }

\begin{document}

\renewcommand{\thefootnote}{\fnsymbol{footnote}}
\setcounter{footnote}{1}


\begin{titlepage}
\pagenumbering{roman}

\vspace*{-1.5cm}
\centerline{\large EUROPEAN ORGANIZATION FOR NUCLEAR RESEARCH (CERN)}
\vspace*{1.5cm}
\noindent
\begin{tabular*}{\linewidth}{lc@{\extracolsep{\fill}}r@{\extracolsep{0pt}}}
\ifthenelse{\boolean{pdflatex}}
{\vspace*{-1.5cm}\mbox{\!\!\!\includegraphics[width=.14\textwidth]{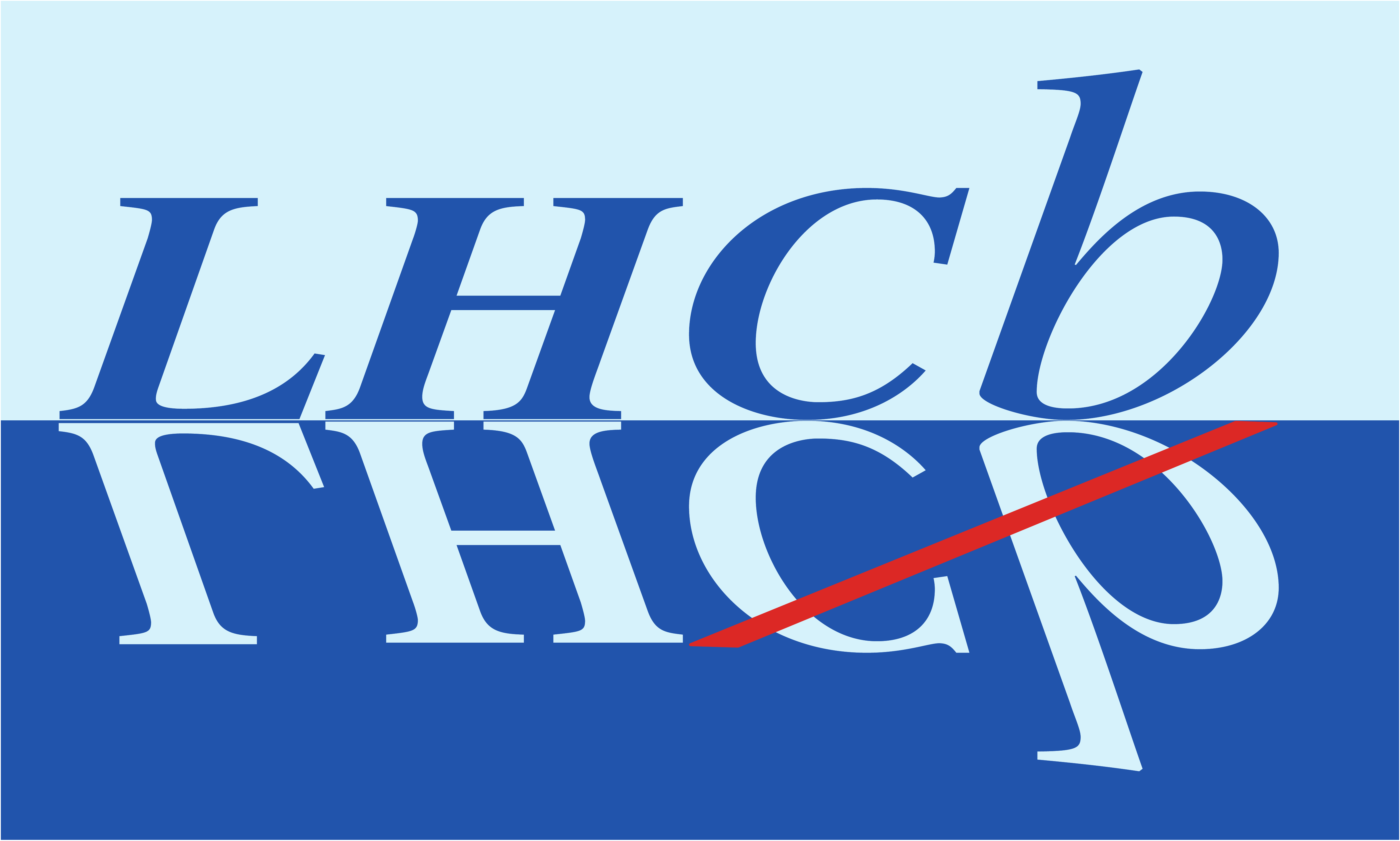}} & &}%
{\vspace*{-1.2cm}\mbox{\!\!\!\includegraphics[width=.12\textwidth]{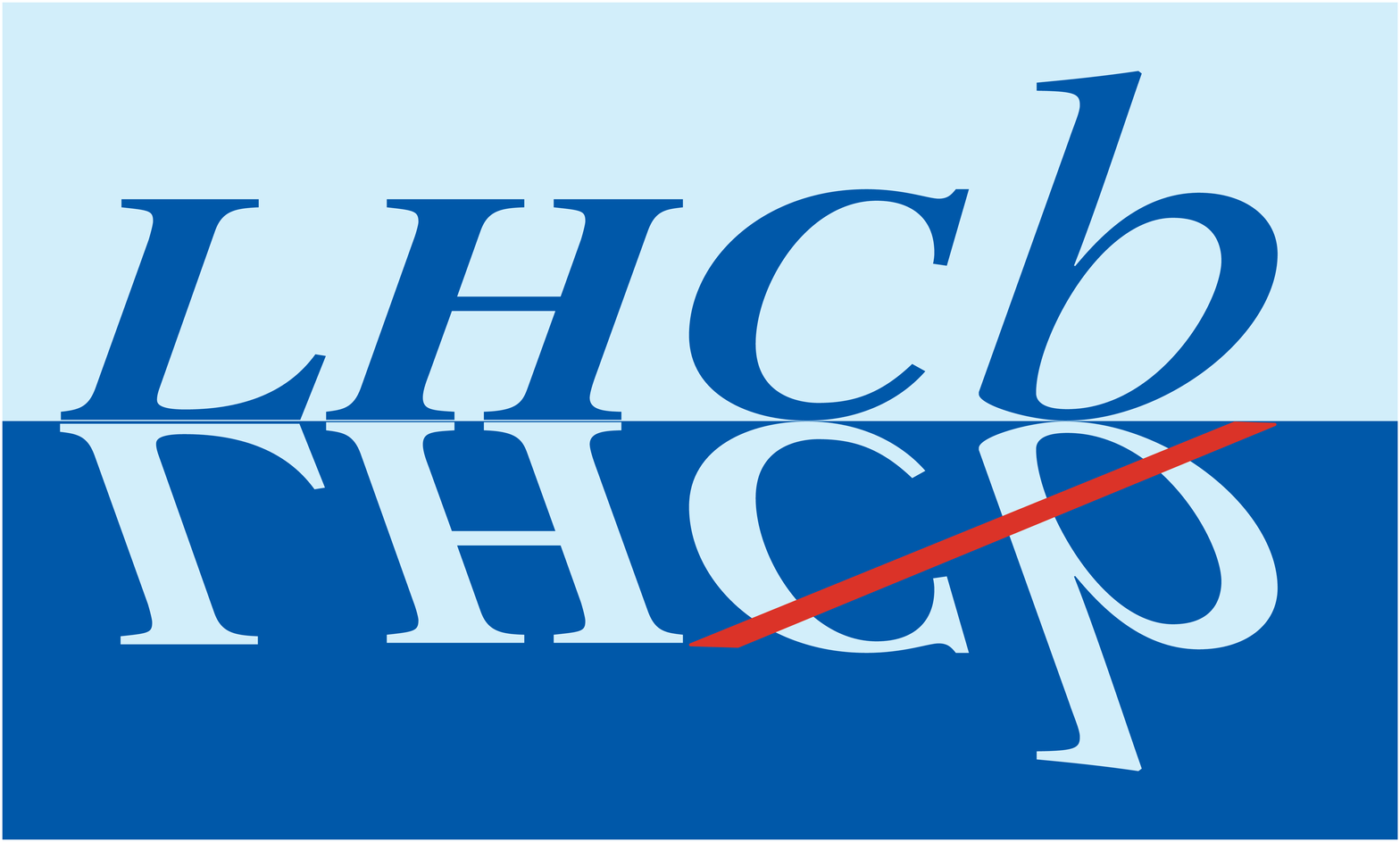}} & &}%
\\
 & & CERN-EP-2019-131 \\  
& & LHCb-PAPER-2019-023 \\  
& &  3 September, 2019 \\ 
 & & \\
\end{tabular*}

\vspace*{4.0cm}

{\normalfont\bfseries\boldmath\huge
\begin{center}
  \papertitle 
\end{center}
}

\vspace*{2.0cm}

\begin{center}
\paperauthors\footnote{Authors are listed at the end of this paper.}
\end{center}

\vspace{\fill}

\begin{abstract}
Using proton\nobreakdash-proton collision data, 
collected with the~\lhcb detector and corresponding to 
1.0, 2.0 and 1.9 \invfb of integrated luminosity at  
the~centre\nobreakdash-of\nobreakdash-mass energies of 7, 8, and 13~\tev, respectively, 
the decay \LbToXPK with \XToJPsipipi  is observed  for the first time. 
The~significance of the~observed signal is in excess of seven standard deviations. 
It is found that $(58\pm15)\%$ of the~decays proceed via 
the~two\nobreakdash-body intermediate state
 ${\Pchi_{\cquark1}\mathrm{(3872)}\Lambda\mathrm{(1520)}}$.
The~branching fraction with respect to that of 
the~\LbToPsiPK decay mode, 
where the~$\psitwos$~meson is reconstructed 
in the~$\jpsi\pip\pim$~final state, is measured to be:
\begin{equation*}
	\dfrac{\BR(\LbToXPK) }{\BR(\LbToPsiPK) } \times \dfrac{\BR(\XToJPsipipi)}{\BR(\PsiToJPsipipi)} =   
	\left(5.4 \pm 1.1 \pm 0.2\right)\times 10^{-2}\,,
\end{equation*} 
where the first uncertainty is statistical and the second is systematic.
\end{abstract}

\vspace*{2.0cm}

\begin{center}
  Published in \href{https://doi.org/10.1007/JHEP09(2019)028}{JHEP (2019) 028}
\end{center}

\vspace{\fill}

{\footnotesize 
\centerline{\copyright~\papercopyright. \href{\paperlicenceurl}{\paperlicence}.}}
\vspace*{2mm}

\end{titlepage}


\newpage
\setcounter{page}{2}
\mbox{~}
%

\cleardoublepage


\renewcommand{\thefootnote}{\arabic{footnote}}
\setcounter{footnote}{0}



\pagestyle{plain} 
\setcounter{page}{1}
\pagenumbering{arabic}


%

\section{Introduction}
\label{sec:Introduction}

The~\chiconex state, also known as \x, was observed in 2003 
by the~\belle collaboration~\cite{Xbelle}
and subsequently confirmed by several other experiments~\cite{Xcdf, Xd0, Xbabar, LHCb-PAPER-2011-034, Xcms,Aaboud:2016vzw}. 
This discovery has attracted much interest 
in exotic charmonium spectroscopy since it was the first observation of an unexpected
charmonium candidate. 
The~mass of the~\chiconex~state has been precisely measured~\cite{LHCb-PAPER-2011-034, cdfX_mass} and the~dipion mass spectrum in the~decay \XToJPsipipi  was also studied~\cite{Xbelle,  Xcms, cdfX_pipi}. The~quantum numbers of the~state were determined to be $\mathrm{J}^{\mathrm{PC}} = 1^{++}$ from measurements performed by the~\lhcb collaboration~\cite{LHCb-PAPER-2015-015}.

 Despite a large amount of experimental information, 
 the nature of the~\chiconex~particle is~still 
 unclear~\cite{XYZ,XYZ_proc}. 
 It has been interpreted as
 a~$\Pchi_{\cquark1}\mathrm{(2P)}$~charmonium
 state~\cite{Achasov:2015oia,Achasov:2017kni},
 molecular state~\cite{Tornqvist:2004qy,Swanson:2003tb,Wong:2003xk},
 tetraquark~\cite{tetraX,Wang:2013vex},
 $\ccbar$g hybrid meson~\cite{hybridX}, 
 vector glueball~\cite{glueballX} or mixed state~\cite{Xmixture2, Xmixture1}. 
 Studies of radiative \chiconex decays~\cite{Aubert:2008ae,LHCb-PAPER-2014-008,Bhardwaj:2011dj} have 
 reduced~the~number 
 of possible interpretations of this state~\cite{Swanson, DongFaeser, FerrettiGalata}. 
 Thus~far, the \chiconex particle has been widely studied 
 in prompt hadroproduction~\cite{Xcdf,Xcms,LHCb-PAPER-2011-034,Aaboud:2016vzw} and  in the weak decays of beauty mesons.
 Several decays of the~$\Lb$~baryon to charmonium 
 have been observed~\cite{LHCb-PAPER-2012-057, LHCb-PAPER-2015-029, LHCb-PAPER-2015-060, LHCb-PAPER-2016-015, LHCb-PAPER-2017-011, LHCb-PAPER-2017-041, LHCb-PAPER-2018-022, PhysRevD.97.072010}. 
 Observing \Lb~decays involving the~\chiconex~state will allow comparison
 of their decay rates to the rates for conventional charmonium states, 
 where, for instance, factorisation and spectator quarks assumptions 
 may lead to different results depending on the nature of the~\chiconex~state.

In this paper the first observation of the \chiconex state in the beauty\nobreakdash-baryon 
decay \mbox{\LbToXPK} is reported. This study is based on data collected with the~\lhcb~detector 
in proton\nobreakdash-proton\,(\proton\proton) collisions corresponding 
to 1.0, 2.0 and 1.9 \invfb~of~integrated luminosity at centre\nobreakdash-of\nobreakdash-mass energies 
of 7, 8 and 13 \tev, respectively. A~measurement of the \LbToXPK branching fraction relative to that of the \mbox{\LbToPsiPK}~decay,
\begin{equation}
	\label{eq:branching}
	R =\dfrac{\BR(\LbToXPK) }{\BR(\LbToPsiPK) } \times \dfrac{\BR(\XToJPsipipi)}{\BR(\PsiToJPsipipi)}\,, 
\end{equation}
 is performed, where the $\chiconex$
and $\psitwos$~mesons are reconstructed in 
the~\JpsiPiPi final state. Throughout this paper the inclusion of charge-conjugated processes is implied.

\section{Detector and simulation}
\label{sec:Detector}

The \lhcb detector~\cite{Alves:2008zz,LHCb-DP-2014-002} is a single-arm forward
spectrometer covering the~pseudorapidity range $2<\eta <5$,
designed for the study of particles containing $\bquark$~or~$\cquark$~quarks. 
The~detector includes a high-precision tracking system consisting of a 
silicon-strip vertex detector surrounding the \proton\proton interaction
region~\cite{LHCb-DP-2014-001}, a large-area silicon-strip detector located
upstream of a dipole magnet with a bending power of about $4{\mathrm{\,Tm}}$,
and three stations of silicon-strip detectors and straw drift tubes~\cite{LHCb-DP-2013-003,LHCb-DP-2017-001} placed downstream of the magnet. 
The tracking system provides a measurement of the momentum of charged particles
with a relative uncertainty that varies from $0.5\%$ at low momentum to $1.0\%$~at~$200 \gevc$. The~minimum distance of a track to a primary vertex\,(PV), the impact parameter\,(IP), 
is~measured with a resolution of $(15+29/\pt)\mum$, where \pt is the component 
of the~momentum transverse to the beam, in\,\gevc. Different types of charged hadrons
are distinguished using information from two ring-imaging Cherenkov detectors\,(RICH)~\cite{LHCb-DP-2012-003}. Photons,~electrons and hadrons are identified 
by a calorimeter system consisting of scintillating-pad and preshower detectors, 
an electromagnetic and a hadronic calorimeter. Muons are~identified by a system 
composed of alternating layers of iron and multiwire proportional chambers~\cite{LHCb-DP-2012-002}.

The online event selection is performed by a trigger~\cite{LHCb-DP-2012-004}, 
which consists of a hardware stage, based on information from the calorimeter and muon systems,
followed by a~software stage, which applies a~full event reconstruction. 
At~the~hardware trigger stage, events are~required to have a muon with high \pt or 
a~pair of opposite\nobreakdash-sign muons  with a~requirement on the product 
of muon transverse momenta, or a hadron, photon or electron with high transverse
energy in the calorimeters. 
The software trigger requires 
two muons of opposite charge forming 
a~good\nobreakdash-quality secondary vertex
with a~mass in excess of 2.7\gevcc,  
or 
a~two-, three- or four\nobreakdash-track secondary vertex 
with  at least one charged particle with a~large \pt 
and inconsistent with originating from any~PV.
For both cases significant displacement of 
the~secondary 
vertex from any primary \proton\proton interaction vertex is required.

Simulated events are used to describe the~signal mass shapes
and compute efficiencies.
In~the~simulation, \proton\proton collisions are generated 
using \pythia~\cite{Sjostrand:2007gs}  with a~specific \lhcb configuration~\cite{LHCb-PROC-2010-056}. 
Decays of unstable particles are described by \evtgen 
package~\cite{Lange:2001uf}, 
in which final-state radiation is generated using \photos~\cite{Golonka:2005pn}. 
The~interaction of the~generated particles with the~detector, 
and its response, are implemented using
the~\geant toolkit~\cite{Allison:2006ve, *Agostinelli:2002hh} 
as described in Ref.~\cite{LHCb-PROC-2011-006}.

\section{Event selection}
\label{sec:Selection}

 The \LbToNR candidate decays are reconstructed using
 \JpsiMuMu decay mode. 
 To~separate signal from background, a loose preselection is applied,
 as done in~Ref.~\cite{LHCb-PAPER-2015-060}, 
 followed by a multivariate classifier based on 
 a~Boosted Decision Tree with gradient boosting\,(BDTG)~\cite{Breiman}.

 Muon, proton, pion and kaon candidates are identified 
 using combined information from the RICH, calorimeter and muon detectors. 
 They are required to have a transverse  momentum larger than 550\mevc for muon 
 and 200\mevc for hadron candidates. To allow for efficient  particle identification,
 kaons and pions are required to have a~momentum between 3.2~and~150\gevc, 
 whilst protons must have a~momentum between 10~and~150\gevc. 
 To~reduce the combinatorial background, 
 only tracks that are inconsistent with originating from any PV are used.
 
Pairs of oppositely charged muons consistent with originating from a common vertex 
are combined to form \JpsiMuMu candidates. 
The~mass of the~pair is required to be
between  $3.0$  and $3.2 \gevcc$. 
 
 To form $\Lb$~candidates, 
 the selected $\jpsi$~candidates 
 are combined with a pair of oppositely charged pions, 
 a~proton and a~negatively charged kaon.  Each~$\Lb$~candidate is associated with
 the~PV that yields the~smallest~$\chisqip$, 
 where \chisqip is defined as the difference in the vertex-fit \chisq of a given PV 
 reconstructed with and without the particle under consideration. 
 The $\chisqip$ value is required to be less than~9. 
 To~improve the $\Lb$\nobreakdash~mass resolution a~kinematic fit~\cite{dtf}  is performed. 
 This~fit constrains the mass of the~$\mumu$~pair to the known mass of the $\jpsi$ meson~\cite{PDG2018}. 
 It~is also required that 
 the~$\Lb$~momentum vector points back 
 to the associated \proton\proton interaction vertex. 
 In~addition, the~measured decay time of 
 the~$\Lb$~candidate, 
 calculated with respect to the~associated PV, 
 is required to be greater than $75 \mum/c$ to suppress poorly reconstructed 
 candidates and background from particles originating from the PV. 
 
 To further suppress cross-feed from the \BzTojpsipipipK decay with 
 a positively charged pion misidentified as a proton, 
 a veto is applied on the $\Lb$~mass, 
 recalculated with a pion mass hypothesis for the proton. 
 A similar veto is applied to suppress \decay{\Bs}{\jpsi\pip\pim\Kp\Km} decays. 
 Any~candidate with a recalculated mass consistent with 
 the known $\Bz$~or $\Bs$~mass is rejected. 
 
 A BDTG is used to further suppress the combinatorial background. 
 It~is~trained on a~simulated sample of \LbToXPK, \mbox{\XToJPsipipi} decays 
 for the signal, while for background the high-mass data sideband is used, defined as 
 \mbox{$m_\JpsiPiPiPK>5640\mevcc$},
 where the regions of $m_{\jpsi\pip\pim}$ populated by  $\mbox{\PsiToJPsipipi}$ and \XToJPsipipi~decays 
 are excluded.
 The~$k$\nobreakdash-fold cross-validation technique~\cite{geisser1993predictive} 
 is used in the BDTG training, in which the candidates are 
 pseudo\nobreakdash-randomly split into $k=23$~samples. 
 The BDTG applied to a particular sample is~trained using 
 all the data from the other $22$, allowing  $\sim95\%$ of the total sample
 to~be used for each training with no need to remove 
 the~candidates used from the final data~set. 
 The~outputs of all multivariate classifiers are consistent. 
 The BDTG is trained on variables related to reconstruction quality,
 kinematics, lifetime of $\Lb$~candidates, 
 the value of \chisq from the kinematic fit described above,
 and the mass of the dipion combination.

 The simulated samples are corrected to better match 
 the kinematic distributions observed in data. 
 The transverse momentum 
 and rapidity distributions and the lifetime of 
 the $\Lb$~baryons
 in simulated samples are adjusted to match those observed in a high-yield 
 low-background sample of \mbox{\LbToJpsiPK} decays.
 Finally, the simulated events are weighted to match the particle identification
efficiencies determined from data using calibration samples
of low-background decays: \DstarPlus, \mbox{\Ksz}, \Dsplus, 
for kaons and pions;
and $\decay{\Lz}{\proton\pim}$~and 
\Lcplus~for protons~\cite{LHCb-DP-2012-003, LHCb-DP-2018-001}. 
The simulated decays of $\Lb$ baryons are produced 
according to a phase-space decay model. The \XToJPsipipi decay proceeds
via the $\jpsi\Prho^0$ S-wave intermediate state~\cite{LHCb-PAPER-2015-015}. 
The simulated \LbToPsiPK decays are corrected to reproduce 
the~$\proton\Km$~mass and $\cos\uptheta_{\proton\Km}$ distributions observed in data, 
where the~helicity angle of the~$\proton\Km$~system, 
$\uptheta_{\proton\Km}$, is 
defined as the~angle between the momentum vectors of the~kaon 
and $\Lb$~baryon in the~$\proton\Km$ rest frame. 
To account for imperfections in the~simulation of
charged particle reconstruction, efficiency corrections obtained using 
data are also applied~\cite{LHCb-DP-2013-002}.
 
 The requirement on the BDTG output $t$ is chosen to maximize
 the Punzi figure of merit $\upvarepsilon_t/(\alpha/2 + \sqrt{B_t})$~\cite{Punzi:2003bu}, 
 where $\upvarepsilon_t$ is the signal efficiency for the~\LbToXPK decay
 obtained from the simulation, $\alpha = 5$ 
 is the~target signal significance
 in units of standard deviations,  
 $B_t$~is the~expected background 
 yield  within narrow mass windows
 centred on the known 
 $\Lb$~and $\chiconex$~masses~\cite{PDG2018}.  
\section{Signal yields and efficiencies}
\label{sec:Sig_eff}

The yields 
for signal and normalization channels 
are determined using a~two\nobreakdash-dimensional
unbinned extended maximum\nobreakdash-likelihood fit
to the \JpsiPiPiPK and \JpsiPiPi masses. 
The probability density function used in the fit consists of four components
to describe the mass spectrum:
\begin{itemize}
	\item[-] a~signal component, describing the true 
	$\decay{\Lb}{\Ppsi_{\Ppi\Ppi}\proton\Km}$ decays, where $\Ppsi_{\Ppi\Ppi}$ denotes either \psitwos or \chiconex  final~states;
	\item[-] a~component describing nonresonant\,(NR) \LbToNR decays 
	with no intermediate $\Ppsi_{\Ppi\Ppi}$~state;
	\item[-] a~component describing random combinations of  
	$\Ppsi_{\Ppi\Ppi}$ with \proton\Km~pairs
	that are not $\Lb$ decay products; 
	\item[-] and a~combinatorial $\jpsi\pip\pim\proton\Km$ component.
\end{itemize}
The templates for the $\Lb$, $\chiconex$ and \psitwos signals 
are described by modified Gaussian functions 
with power-law tails on both sides~\cite{Skwarnicki:1986xj}. 
The tail parameters are fixed to values obtained from simulation,
while  the~peak positions of the Gaussian functions are free to vary 
in the~fit.
The~mass resolution of the $\psitwos$~meson is allowed to vary in the~fit,
while that of the~$\chiconex$~signal,
due to its lower yield, is fixed to
the~value determined from simulation
and corrected by the~data\nobreakdash-simulation ratio
of the~mass resolutions for the~$\psitwos$ meson.
The~\mbox{$\decay{\Lb}{\Ppsi_{\Ppi\Ppi}\proton\Km}$}~component 
is described by the product of the~$\Lb$ and $\Ppsi_{\Ppi\Ppi}$~signal templates,
\mbox{$S_{\Lb}(m_{\jpsi\pip\pim\proton\Km})\times S_{\Ppsi_{\Ppi\Ppi}}(m_{\jpsi\pip\pim})$}. 
The~NR~$\decay{\Lb}{\jpsi\pip\pim\proton\Km}$~component is described 
by the product of the~$\Lb$~signal template, an~exponential function
and a~first\nobreakdash-order polynomial function,
\mbox{$S_{\Lb}(m_{\jpsi\pip\pim\proton\Km})\times E(m_{\jpsi\pip\pim})\times P_1(m_{\jpsi\pip\pim})$},
while the~$\Ppsi_{\Ppi\Ppi}\proton\Km$~component
is parametrized as the product of the~$\Ppsi_{\Ppi\Ppi}$~signal template and 
an~exponential function, $S_{\Ppsi_{\Ppi\Ppi}}(m_{\jpsi\pip\pim})\times 
E(m_{\jpsi\pip\pim\proton\Km})$.
The~combinatorial background is modelled by the function
\begin{equation}
\begin{aligned}
	f(m_{\jpsi\pip\pim\proton\Km}, 
	m_{\jpsi\pip\pim}) & = 
	E(m_{\jpsi\pip\pim\proton\Km})
	\times   
	\Phi_{3,5}(m_{\jpsi\pip\pim}) 
	\\ 
	& \times  
	P_3(m_{\jpsi\pip\pim\proton\Km}, 
	m_{\jpsi\pip\pim}),
\end{aligned}
\end{equation}
where $\Phi_{3,5}(m_{\jpsi\pip\pim})$ is a~three\nobreakdash-body\,(\jpsi\pip\pim)
phase space function of the five-body \Lb~decay~\cite{Byckling}, and $P_3$ is a~two\nobreakdash-dimensional positive third-order polynomial function in Bernstein form.

\begin{figure}[t]
	\setlength{\unitlength}{1mm}
	\centering
	\begin{picture}(150,130)
	
	\put( 0,65){\includegraphics*[width=75mm,height=60mm]{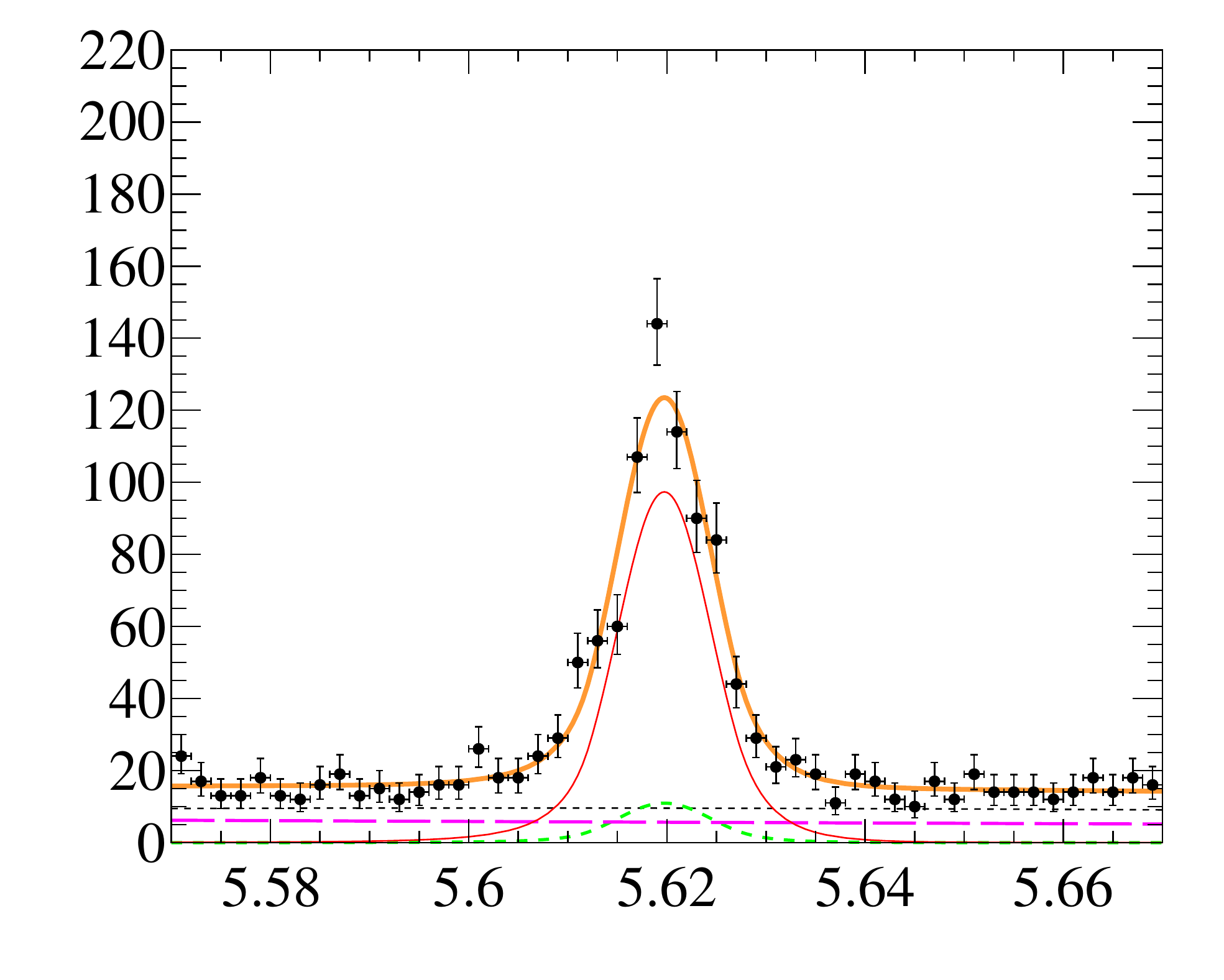}}
	\put(75,65){\includegraphics*[width=75mm,height=60mm]{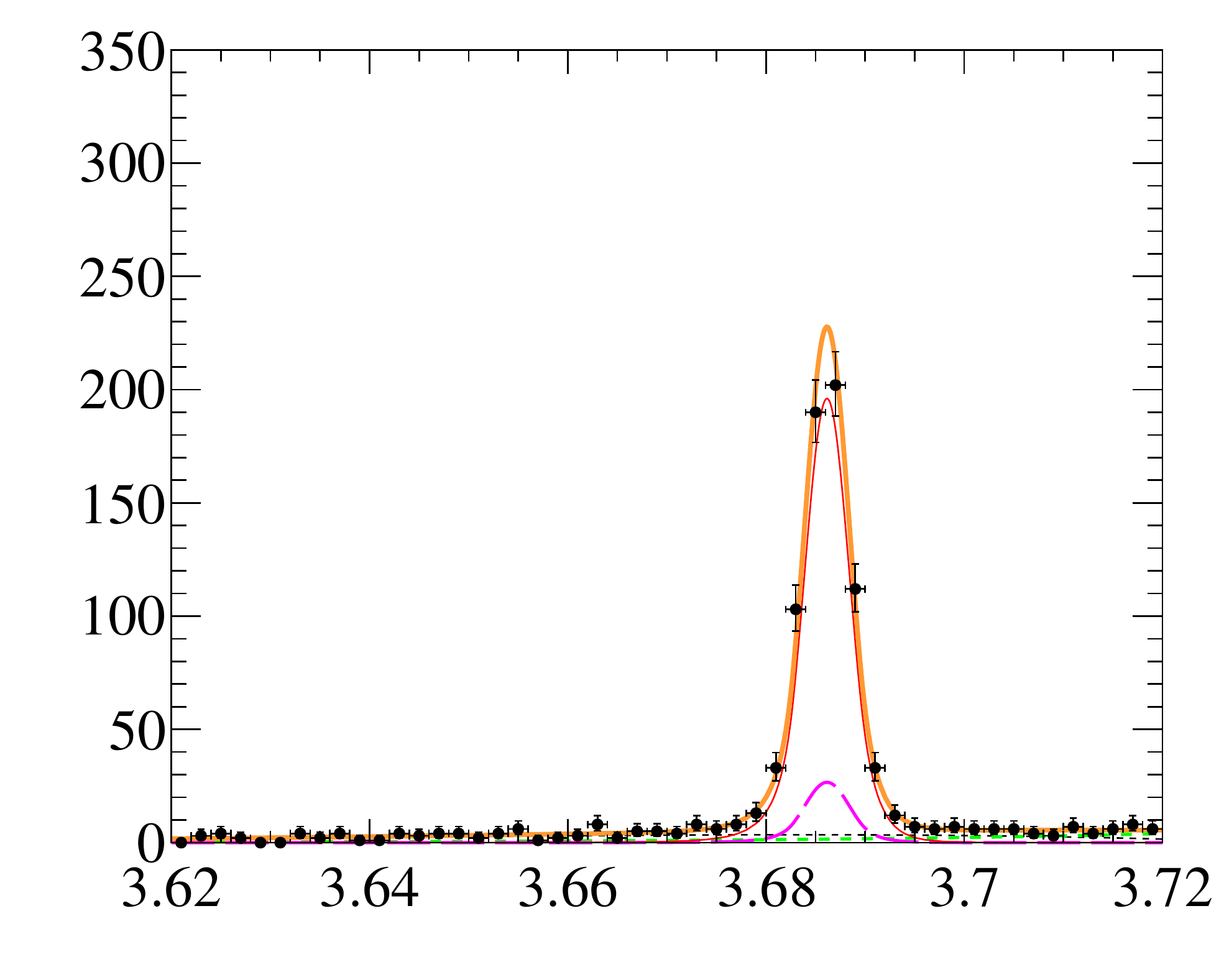}}
	
        \put(0,0){\includegraphics*[width=75mm,height=60mm]{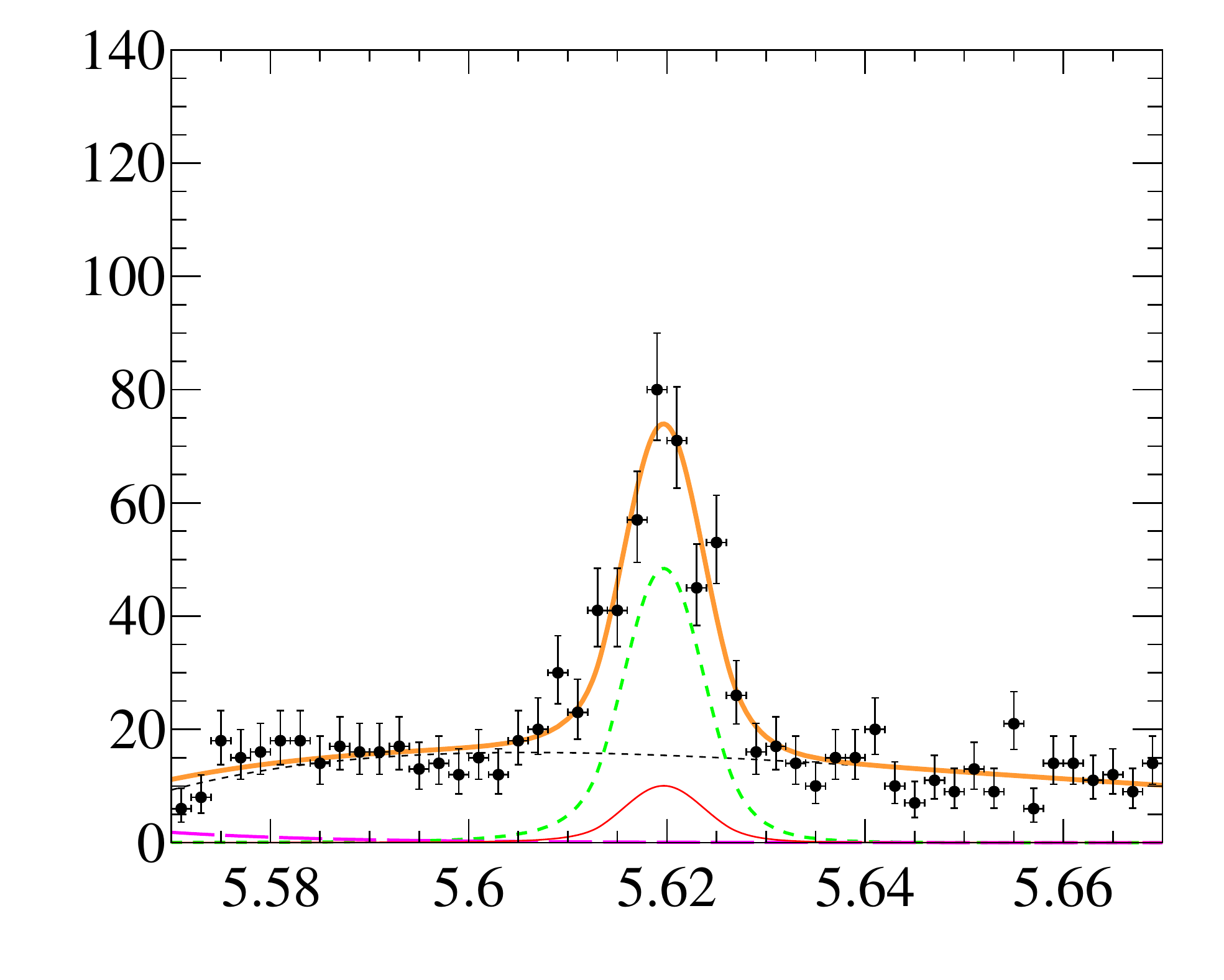}}
	\put(75,0){\includegraphics*[width=75mm,height=60mm]{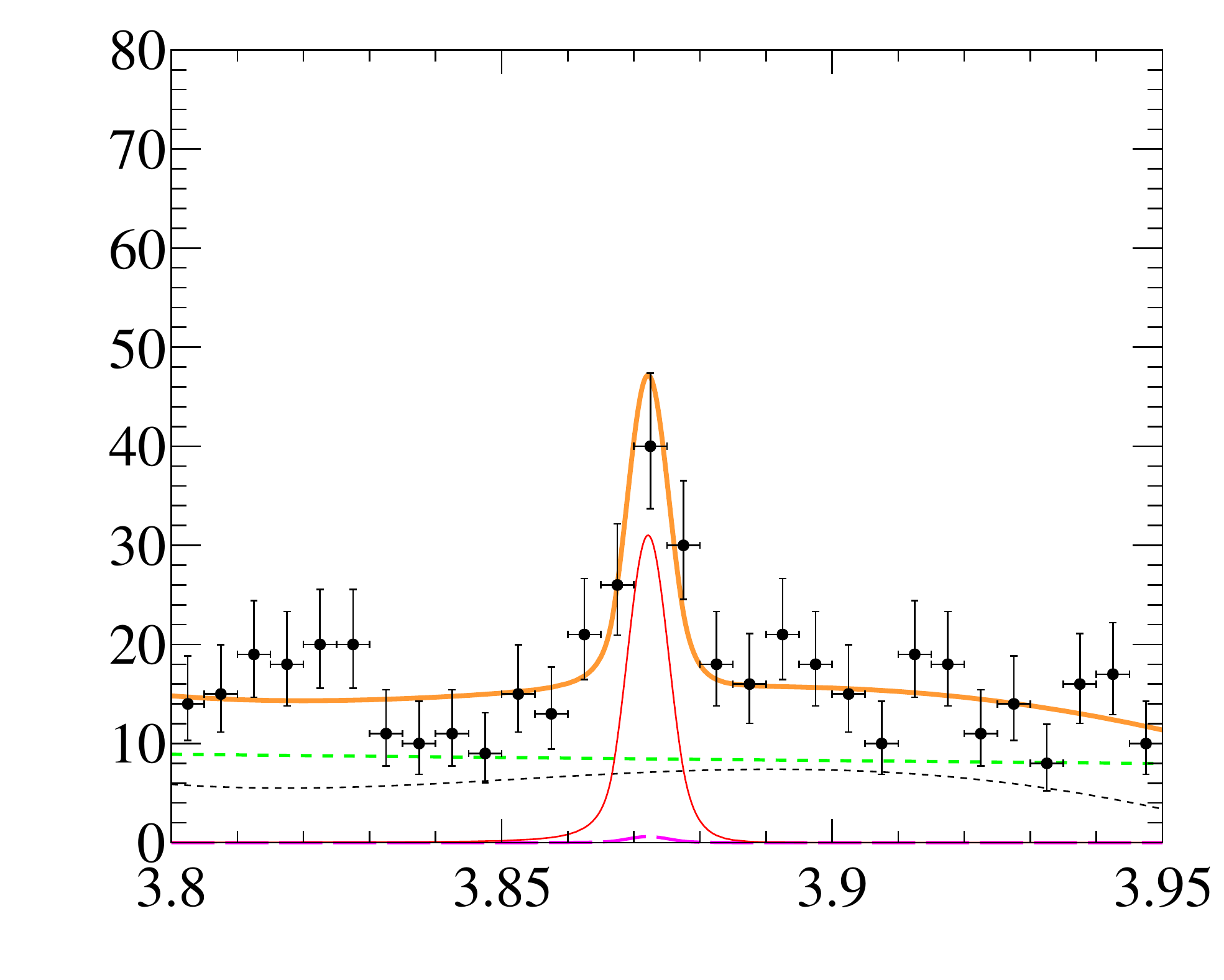}}

	\put(  0,85){\begin{sideways}\small{Candidates/$(2\mevcc)$}\end{sideways}}
	\put( 75,85){\begin{sideways}\small{Candidates/$(2\mevcc)$}\end{sideways}}
	\put(30 ,65){$m_\JpsiPiPiPK$}
	\put(105,65){$m_\JpsiPiPi$}
	\put( 57,65){$\left[\!\gevcc\right]$}
	\put(132,65){$\left[\!\gevcc\right]$}
	
	\put(  0,20){\begin{sideways}\small{Candidates/$(2\mevcc)$}\end{sideways}}
	\put( 75,20){\begin{sideways}\small{Candidates/$(5\mevcc)$}\end{sideways}}
	\put(30 , 0){$m_\JpsiPiPiPK$}
	\put(105, 0){$m_\JpsiPiPi$}
	\put( 57, 0){$\left[\!\gevcc\right]$}
	\put(132, 0){$\left[\!\gevcc\right]$}
	
	\put( 57,111){\lhcb}
	\put(132,111){\lhcb}
	
	\put( 57, 46){\lhcb}
	\put(132, 46){\lhcb}


	\put( 15,116){\scriptsize$3.62<m_{\jpsi\pip\pim}<3.72\gevcc$}
	\put( 15, 51){\scriptsize$3.80<m_{\jpsi\pip\pim}<3.95\gevcc$}
	
	\put( 90,116){\scriptsize$5.61<m_{\jpsi\pip\pim\proton\Km}<5.63\gevcc$}
    \put( 90, 51){\scriptsize$5.61<m_{\jpsi\pip\pim\proton\Km}<5.63\gevcc$}
		

	\put(88,112){\color[rgb]{1,0,0}     {\rule{5mm}{1.0pt}}}
	\put(88,107){\color{green}          {\hdashrule[0.0ex][x]{5mm}{1.0pt}{1.2mm 0.3mm} } }
	\put(88,102){\color{magenta}        {\hdashrule[0.0ex][x]{5mm}{1.0pt}{2.0mm 0.3mm} } }
	\put(88,97){\color{black}          {\hdashrule[0.0ex][x]{5mm}{1.0pt}{0.5mm 0.3mm} } }
	\put(88,92){\color[RGB]{255,153,51} {\rule{5mm}{2.0pt}}}
	
	\put( 95,111.5){\scriptsize{\decay{\Lb}{\Ppsi_{\Ppi\Ppi}\proton\Km}}}
	\put( 95,106.5){\scriptsize{\LbToNR\,(NR)}}
	\put( 95,101.5){\scriptsize{$\Ppsi_{\Ppi\Ppi}\proton\Km$}}
	\put( 95,96.5){\scriptsize{combinatorial bkg.}}
	\put( 95,91.5){\scriptsize{total}}
	\end{picture}
	\caption {\small Projection of the two-dimensional distributions of 
	(left)~\JpsiPiPiPK and (right)~\JpsiPiPi masses for the (top)~\LbToPsiPK 
	and (bottom)~$\mbox{\decay{\Lb}{\Pchi_{\cquark1}\mathrm{(3872)}\proton\Km}}$~candidates.}
	\label{fig:signal_norm}
\end{figure}

Projections of the two-dimensional fits to the~\JpsiPiPiPK and \JpsiPiPi mass distributions for the intervals of \mbox{$3.62< m_{\jpsi\pip\pim} <3.72\gevcc$}   and \mbox{$3.80<m_{\jpsi\pip\pim} <3.95\gevcc$}
are shown in Fig.~\ref{fig:signal_norm}. The signal yields are determined to be 
 $610\pm30$ and $55 \pm 11$ for the \mbox{\LbToPsiPK} and \LbToXPK decay modes, respectively.
The statistical significance of the observed 
$\decay{\Lb}{\Pchi_{\cquark1}\mathrm{(3872)}\proton\Km}$
signal is estimated to be $7.2\sigma$ using Wilks' theorem~\cite{Wilks:1938dza} 
and confirmed by simulating a large number of pseudoexperiments 
according to the background distributions observed in data.

\begin{figure}[t]
	\setlength{\unitlength}{1mm}
	\centering
	\begin{picture}(150,120)
	
	\put( 0,0){\includegraphics*[width=150mm,height=120mm]{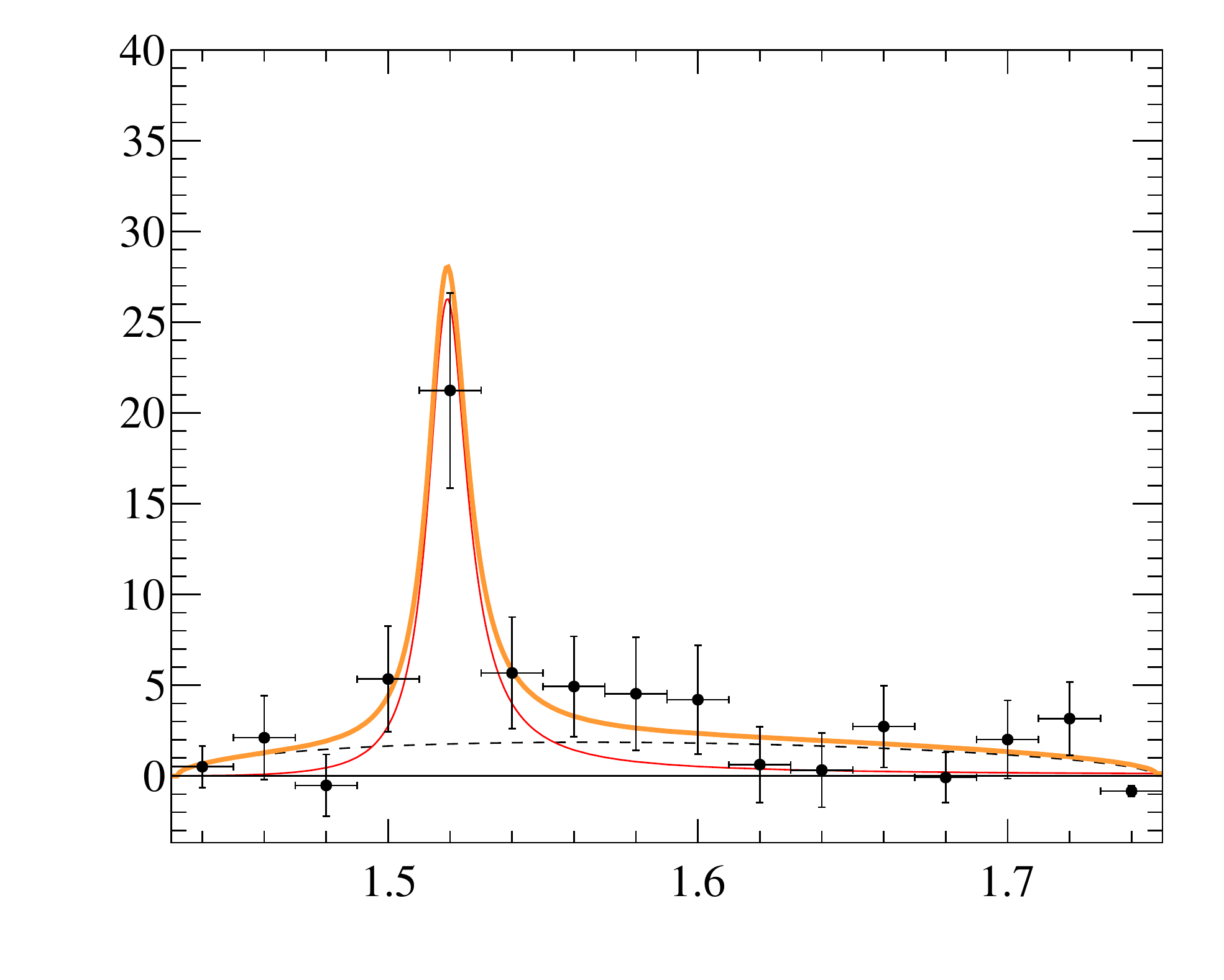}}

	\put(0,45){\begin{sideways}\Large{$\rm{N_{\LbToXPK}}/(20\mevcc)$}\end{sideways}}

	\put(75,3){\Large{$m_{\proton\Km}$}}
	\put(122,3){\Large{$\left[\!\gevcc\right]$}}
	
	\put( 120,100){\Large{\lhcb}}

	
	\put(70,90){\color[rgb]{1,0,0}       {\rule{10mm}{1.0pt}}}
	\put(70,80){\color{black}    		 {\hdashrule[0.0ex][x]{10mm}{1.0pt}{1.5mm 0.5mm} } }
	\put(70,70){\color[RGB]{255,153,51}  {\rule{10mm}{3.0pt}}}

	\put(83,88.5){\large{$\decay{\Lb}{\Pchi_{\cquark1}\mathrm{(3872)}\Lambdares}$}}
	\put(83,78.5){\large{$\decay{\Lb}{\chiconex(\proton\Km)_{\rm{NR}}}$}}
	\put(83,69.5){\large{total}}
 
	\end{picture}
	\caption {\small Background-subtracted mass distribution for the \proton\Km~system in \mbox{\LbToXPK}~decays with fit results in  the range \mbox{$1.43 < m_{\proton\Km} < 1.75 \gevcc$} superimposed. The~background subtraction is performed using the~sPlot~technique~\cite{Pivk:2004ty}.}
	\label{fig:res_mpk}
\end{figure}

The background-subtracted \proton\Km~mass spectrum~\cite{Pivk:2004ty}
for the signal 
channel 
is shown in Fig.~\ref{fig:res_mpk}. 
The distribution exhibits a clear peak 
associated with the~\Lambdares state.
From~this~distribution the~fraction of 
two\nobreakdash-body \LbToLamst decays is determined using 
an unbinned maximum-likelihood fit, which includes two components. 
The~first component corresponds to the \LbToLamst  decay and is described 
with a relativistic P-wave Breit$-$Wigner function. 
The second component corresponds to the~nonresonant decay 
\mbox{\LbToXPK} and is modelled by
\begin{equation}
	B(m_{\proton\Km}) = \Phi_{2,3}(m_{\proton\Km}) \times P_1(m_{\proton\Km}),
\end{equation}
\noindent where $\Phi_{2,3}(m_{\proton\Km})$ is 
a~two-body\,(\proton\Km) 
phase space function of the three-body decay of the $\Lb$ baryon
and $P_1(m_{\proton\Km})$~a first-order polynomial function.  
The peak position and the natural width are 
constrained to the known values for the $\Lambdares$~resonance~\cite{PDG2018}. 
The~fraction of \mbox{\LbToLamst} decays obtained from the fit is $(58 \pm 15)\%$, 
where the~uncertainty is statistical only.

The ratio $R$ defined in Eq.~\eqref{eq:branching} is obtained as
\begin{equation}
	R = 	\dfrac{N_{\chiconex\proton\Km}}{N_{\psitwos\proton\Km}} \times \dfrac{\upvarepsilon_{\psitwos\proton\Km}}{\upvarepsilon_{\chiconex\proton\Km}}\,,
\end{equation}
where $N$ represents the measured yield and $\upvarepsilon$ 
denotes the efficiency of the corresponding decay. 
The efficiency is defined as the product of the geometric acceptance and 
the~detection, reconstruction, selection and trigger efficiencies. 
All efficiencies are determined using corrected simulated samples.

The efficiencies are determined separately for each data-taking period
and are combined according to the corresponding integrated luminosities~\cite{LHCb-PAPER-2014-047} 
for each period and the known 
cross-section of $\bquark$\nobreakdash-hadron production 
in the \lhcb acceptance~\cite{LHCb-PAPER-2010-002, LHCb-PAPER-2011-003, LHCb-PAPER-2013-016, LHCb-PAPER-2015-037, LHCb-PAPER-2016-031}. 
The~ratio of the~efficiency of the~normalization channel 
to that of the~signal channel is determined~to~be
\begin{equation}
\dfrac{\upvarepsilon_{\psitwos\proton\Km}}{\upvarepsilon_{\chiconex\proton\Km}} = 0.6065 \pm 0.0035\,,
\label{eq:eff_ratio}
\end{equation}
where only the~uncertainty that arises from the~sizes of the~simulated samples is given. 
\mbox{Additional} sources  of uncertainty are discussed in the~following section. 
The~ratio of efficiencies differs from unity
mainly due to different dipion mass spectra in 
the~\mbox{$\decay{\Pchi_{\cquark1}\mathrm{(3872)}}{\jpsi\pip\pim}$}
and $\decay{\Ppsi\mathrm{(2S)}}{\jpsi\pip\pim}$~decays.

\section{Systematic uncertainties}
\label{sec:Systematics}

Since the signal and normalization decay channels have similar kinematics
and topologies, a large part of systematic uncertainties cancel in the ratio $R$. 
The remaining contributions to the systematic uncertainty are 
listed in Table~\ref{tab:systematics} and discussed below.

To estimate the systematic uncertainty related to the fit model,
pseudoexperiments are generated according to the mass shapes 
obtained from the data fit. Each pseudoexperiment is then fitted
with the baseline fit and alternative signal models and the ratio $R$ is computed.
A~generalized Student's $t$-distribution~\cite{Jackman}, 
an~Apollonios function~\cite{Santos:2013gra} and 
a~modified Novosibirsk function~\cite{PhysRevD.84.112007} 
are used as alternative models for the signal component. 
The~maximum relative bias found for the ratio $R$ is 2\%, 
which is assigned as a relative systematic uncertainty.

The simulated \LbToPsiPK decays are corrected to reproduce 
the~$\proton\Km$~mass and $\cos\uptheta_{\proton\Km}$ 
distributions observed in data. 
The~uncertainty associated with this~correction procedure and 
related to the imperfect knowledge of 
the~\LbToPsiPK decay model is estimated by varying 
the reference kinematic 
$m_{\proton\Km}$ and 
$\cos\uptheta_{\proton\Km}$~distributions
within their uncertainties.
It~causes a~negligible change of 
the~efficiency~$\upvarepsilon_{\psitwos\proton\Km}$. 
A~similar procedure applied to the \LbToXPK channel 
leads to a systematic uncertainty of $2\%$ on the efficiency $\upvarepsilon_{\chiconex\proton\Km}$.

An additional uncertainty arises from the 
differences between data and simulation, in particular
those affecting the efficiency for the reconstruction of charged-particle tracks.
The small difference in the track-finding efficiency 
between data and simulation is corrected 
using data~\cite{LHCb-DP-2013-002}. 
The uncertainties in these correction factors together
with the uncertainties in the hadron-identification efficiencies, 
related to the finite size of the calibration samples~\cite{LHCb-DP-2012-003, LHCb-DP-2018-001},
are propagated to the ratio of total efficiencies using pseudoexperiments. 
This results in a systematic uncertainty of $0.4\%$ 
associated with track reconstruction and hadron identification.

\begin{table}[t]
	\centering
	\caption{\small
		Relative systematic uncertainties for the~ratio of branching fractions.
	}\label{tab:systematics}
	\vspace*{3mm}
	\begin{tabular*}{0.80\textwidth}{@{\hspace{3mm}}l@{\extracolsep{\fill}}c@{\hspace{3mm}}}
		Source              & Uncertainty~$\left[\%\right]$
		\\[1mm]
		\hline
		\\[-3mm]
		Fit model                           & 2.0        \\
		Decay model of the \LbToXPK channel        & 2.0        \\
		Track reconstruction and hadron identification & 0.4                   \\
		Trigger                             & 1.7          \\
		Selection criteria                  & 1.0          \\
		Size of the~simulated samples      & 0.6             
		\\[1mm]
		\hline
		\\[-3mm]
		Sum in quadrature                              & 3.5
	\end{tabular*}
\end{table}

To~probe a~possible mismodelling of the~trigger efficiency, 
the ratio of efficiencies is calculated for 
various subsamples, matched to different trigger objects, 
namely dimuon vertex,
high-\pt~$\mumu$~pair, 
two-, three- and four\nobreakdash-track secondary vertex, \etc
The~small difference of 1.7\%
in the ratio of trigger efficiencies
between different subsamples is taken as 
systematic uncertainty due the trigger efficiency estimation. 
Another source of uncertainty is the potential disagreement
between data and simulation in the estimation of efficiencies, 
due to effects not considered above. This is studied by varying 
the selection criteria in ranges that lead to
as much as $\pm 20\%$ change in the~measured signal yields. 
The~stability is tested by comparing the efficiency-corrected
yields within these variations. The resulting variations in the 
efficiency-corrected yields do not exceed~$1\%$, 
which is taken as a corresponding systematic uncertainty~\cite{LHCb-PAPER-2018-022}.
The $0.6\%$ relative uncertainty in the ratio of efficiencies
from Eq.~\eqref{eq:eff_ratio} is assigned as a~systematic uncertainty 
due to the finite size of the simulated samples.

The systematic uncertainty on the fraction of 
$\Lb$~baryons decaying to the~$\Lambdares$~resonance 
is calculated by varying the parameters of the resonant and nonresonant components in the~fit and found to be negligible with respect to the~statistical uncertainty.
\section{Results and summary}
\label{sec:Results}

The decay \LbToXPK with \XToJPsipipi is observed 
using data  collected with the~\lhcb detector in
proton-proton collisions corresponding to  1.0, 2.0 and 1.9 \invfb of integrated luminosity 
at the centre-of-mass energies of 7, 8, and 13~\tev, respectively. 
The observed yield of \LbToXPK decays is  $55 \pm 11$ 
with a~statistical significance in excess of seven standard deviations.
It is found that $(58\pm15)\%$ of the~decays proceed via the two\nobreakdash-body
 ${\Pchi_{\cquark1}\mathrm{(3872)}\Lambda\mathrm{(1520)}}$ intermediate state.

Using the \LbToPsiPK, \PsiToJPsipipi decay as a normalization channel, 
the~ratio of the branching fractions is measured to be
\begin{equation*}
	R =\dfrac{\BR(\LbToXPK) }{\BR(\LbToPsiPK) } \times \dfrac{\BR(\XToJPsipipi)}{\BR(\PsiToJPsipipi)} = 
	\left(5.4 \pm 1.1 \pm 0.2\right)\times 10^{-2}\,, 
\end{equation*}
where the first uncertainty is statistical and the second is systematic.

 Using the values of  $\BR(\LbToPsiPK)$ and $\BR(\PsiToJPsipipi)$ taken from Ref.~\cite{PDG2018} 
 the product of branching fractions of interest is calculated to be
\begin{equation*}
	\BR(\LbToXPK) \times \BR(\XToJPsipipi) = 
	 \left(1.2 \pm 0.3 \pm0.2 \right) \times 10^{-6}\,,
\end{equation*}
where the first uncertainty is statistical and the second is systematic,
including the~uncertainties on the branching fractions $\BR(\LbToPsiPK)$ and $\BR(\PsiToJPsipipi)$.

\section*{Acknowledgements}
%
%
\noindent We express our gratitude to our colleagues in the CERN
accelerator departments for the~excellent performance of the LHC. 
We~thank the technical and administrative staff at the LHCb
institutes.
We~acknowledge support from CERN and from the national agencies:
CAPES, CNPq, FAPERJ and FINEP\,(Brazil); 
MOST and NSFC\,(China); 
CNRS/IN2P3\,(France); 
BMBF, DFG and MPG\,(Germany); 
INFN\,(Italy); 
NWO\,(Netherlands); 
MNiSW and NCN\,(Poland); 
MEN/IFA\,(Romania); 
MSHE\,(Russia); 
MinECo\,(Spain); 
SNSF and SER\,(Switzerland); 
NASU\,(Ukraine); 
STFC\,(United Kingdom); 
DOE NP and NSF\,(USA).
We acknowledge the computing resources that are provided by CERN, 
IN2P3\,(France), 
KIT and DESY\,(Germany), 
INFN\,(Italy), 
SURF\,(Netherlands),
PIC\,(Spain), 
GridPP\,(United Kingdom), 
RRCKI and Yandex LLC\,(Russia), 
CSCS\,(Switzerland), 
IFIN\nobreakdash-HH\,(Romania), 
CBPF\,(Brazil),
PL\nobreakdash-GRID\,(Poland) and
OSC\,(USA).
We~are indebted to the communities behind the~multiple open-source
software packages on which we depend.
Individual groups or members have received support from
AvH Foundation (Germany);
EPLANET, Marie Sk\l{}odowska\nobreakdash-Curie Actions and ERC\,(European Union);
ANR, Labex P2IO and OCEVU, and 
R\'{e}gion Auvergne\nobreakdash-Rh\^{o}ne\nobreakdash-Alpes\,(France);
Key Research Program of Frontier Sciences of CAS, CAS PIFI, and 
the Thousand Talents Program\,(China);
RFBR, RSF and Yandex~LLC\.(Russia);
GVA, XuntaGal and GENCAT\,(Spain);
the Royal Society
and the~Leverhulme Trust\,(United Kingdom).

\addcontentsline{toc}{section}{References}
\bibliographystyle{LHCb}
\bibliography{main,standard,LHCb-PAPER,LHCb-CONF,LHCb-DP,LHCb-TDR}
 
\newpage

\centerline
{\large\bf LHCb collaboration}
\begin
{flushleft}
\small
R.~Aaij$^{29}$,
C.~Abell{\'a}n~Beteta$^{46}$,
T.~Ackernley$^{56}$,
B.~Adeva$^{43}$,
M.~Adinolfi$^{50}$,
C.A.~Aidala$^{78}$,
S.~Aiola$^{23}$,
Z.~Ajaltouni$^{7}$,
S.~Akar$^{61}$,
P.~Albicocco$^{20}$
% ...

{\footnotesize \it

% ...
$ ^{7}$Universit{\'e} Clermont Auvergne, CNRS/IN2P3, LPC, Clermont-Ferrand, France\\
% ...
$ ^{20}$INFN Laboratori Nazionali di Frascati, Frascati, Italy\\
% ...
$ ^{23}$INFN Sezione di Milano, Milano, Italy\\
% ...
$ ^{29}$Nikhef National Institute for Subatomic Physics, Amsterdam, Netherlands\\
% ...
$ ^{43}$Instituto Galego de F{\'\i}sica de Altas Enerx{\'\i}as (IGFAE), Universidade de Santiago de Compostela, Santiago de Compostela, Spain\\
% ...
$ ^{46}$Physik-Institut, Universit{\"a}t Z{\"u}rich, Z{\"u}rich, Switzerland\\
% ...
$ ^{50}$H.H. Wills Physics Laboratory, University of Bristol, Bristol, United Kingdom\\
% ...
$ ^{56}$Oliver Lodge Laboratory, University of Liverpool, Liverpool, United Kingdom\\
% ...
$ ^{61}$University of Cincinnati, Cincinnati, OH, United States\\
% ...
$ ^{64}$Syracuse University, Syracuse, NY, United States\\
% ...
$ ^{78}$University of Michigan, Ann Arbor, United States, associated to $^{64}$\\
\bigskip
% ...
}
\end{flushleft}

\end{document}





\section*{Supplementary material for LHCb-PAPER-2019-023}

This appendix contains supplementary material that will be posted
on the public CDS record but will not appear in the paper.

\subsection*{Comparison of the branching fraction ratios
  with analogous results 
  obtained in B-meson decays}
  
The result of this analysis is consistent with analogous
measurements in neutral and charged B-meson decays 
to the \jpsi, \chicone, and \psitwos final~states~\cite{PDG2018}.
The~ratios are defined as
\begin{eqnarray*}
\label{eq:rat_jpsi}
R_\jpsi & \equiv & \dfrac{\BR(\rm{X_\bquark\to\chiconex + X})}
{\BR(\rm{X_\bquark\to\jpsi} + X)} \times  \BR(\XToJPsipipi)\,,
\\
R_\chicone & \equiv & \dfrac{\BR(\rm{X_\bquark\to\chiconex + X})}
{\BR(\rm{X_\bquark\to\chicone} + X)} \times  \BR(\XToJPsipipi)\,,
\\
R_\psitwos & \equiv & \dfrac{\BR(\rm{X_\bquark\to\chiconex + X})}
{\BR(\rm{X_\bquark\to\psitwos} + X)} \times
\dfrac{\BR(\XToJPsipipi)}{\BR(\PsiToJPsipipi)}\,, 
\end{eqnarray*}
where $\rm{X_\bquark}$~denotes a~beauty~hadron, 
and $\mathrm{X}$~denotes $\Kstarz, \Kz,\pip,\Kz\pip$ or $\proton\Km$. 

\begin{figure}[hbt]
  \setlength{\unitlength}{1mm}
  \centering
  \begin{picture}(150,120)
    \put( -5,5){\includegraphics*[width=65mm,height=120mm]{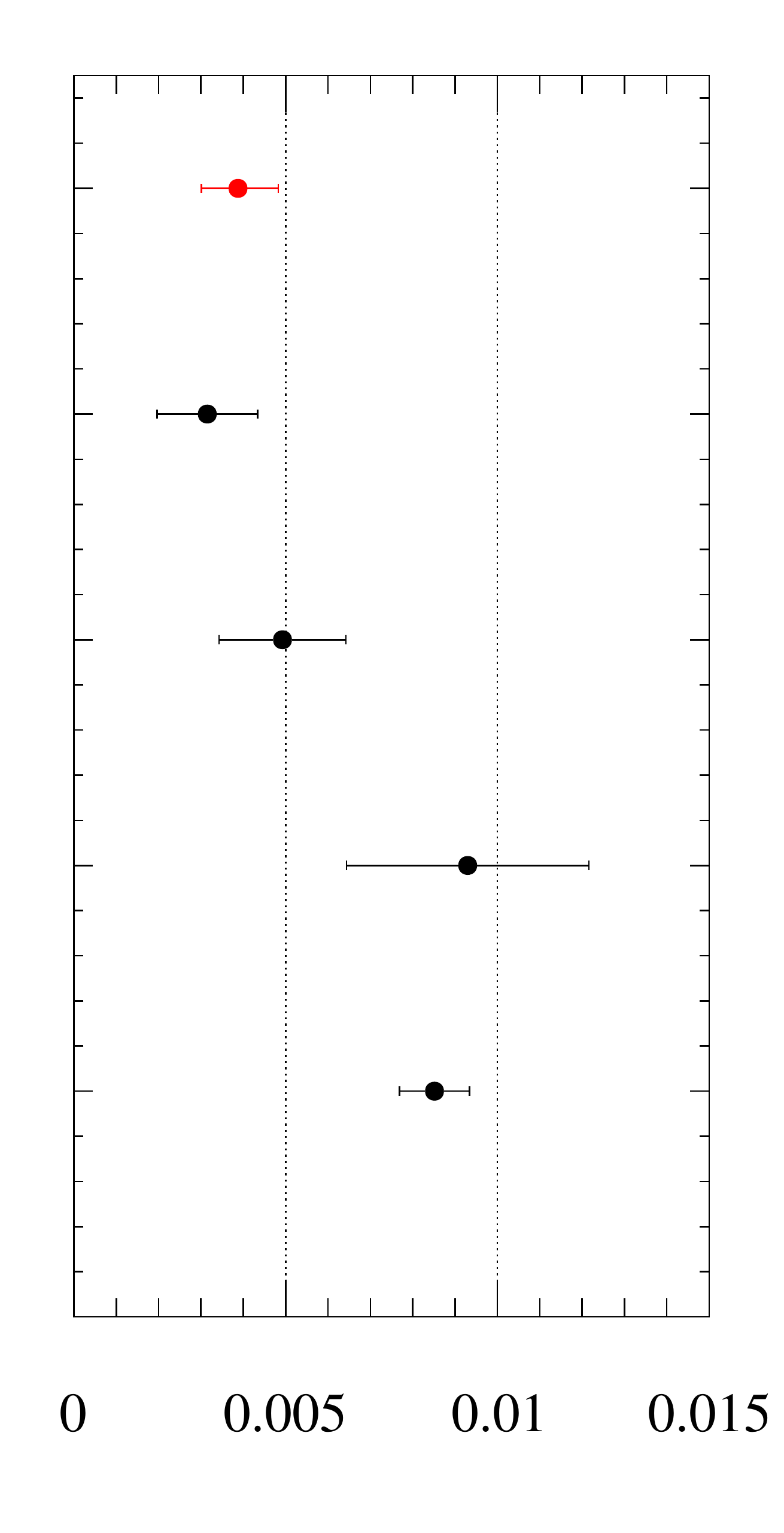}}
    \put(90,5){\includegraphics*[width=65mm,height=120mm]{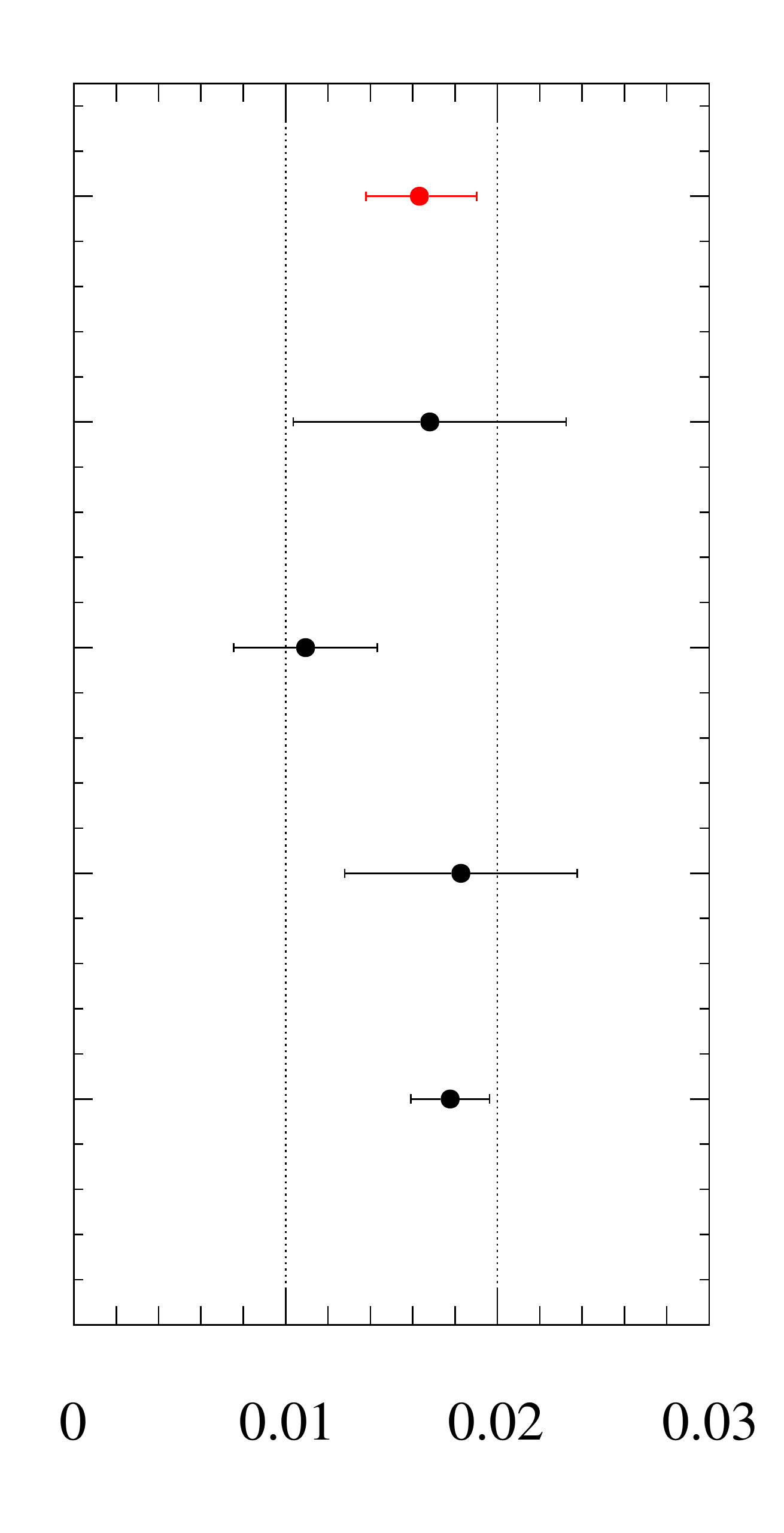}}
    \put(62,109){\large $\Lb\to\Ppsi\proton\Km$}
    \put(62, 91){\large $\Bz\to\Ppsi\Kstarz$}
    \put(62, 73){\large $\Bz\to\Ppsi\Kz$}
    \put(62, 55){\large $\Bu\to\Ppsi\Kz\pip$}
    \put(62, 37){\large $\Bu\to\Ppsi\pip$}
    \put( 22,5)  {\large$R_\jpsi$}
    \put(120,5)  {\large$R_\chicone$}
  \end{picture}
  \caption {\small
  Comparison of the branching fraction ratios
  measured in this analysis (red dot) with analogous results 
  obtained in neutral and charged B-meson decays~\cite{PDG2018} to 
  (left)~the \jpsi 
  and (right)~\chicone final-state. 
  The symbol \Ppsi in decay chains denotes a~$\jpsi$, 
  $\chicone$ or $\chiconex$~meson.}
  \label{fig:ratios}
\end{figure}

\begin{figure}[ht]
  \setlength{\unitlength}{1mm}
  \centering
  \begin{picture}(150,120)
    \put( 60,5){\includegraphics*[width=65mm,height=120mm]{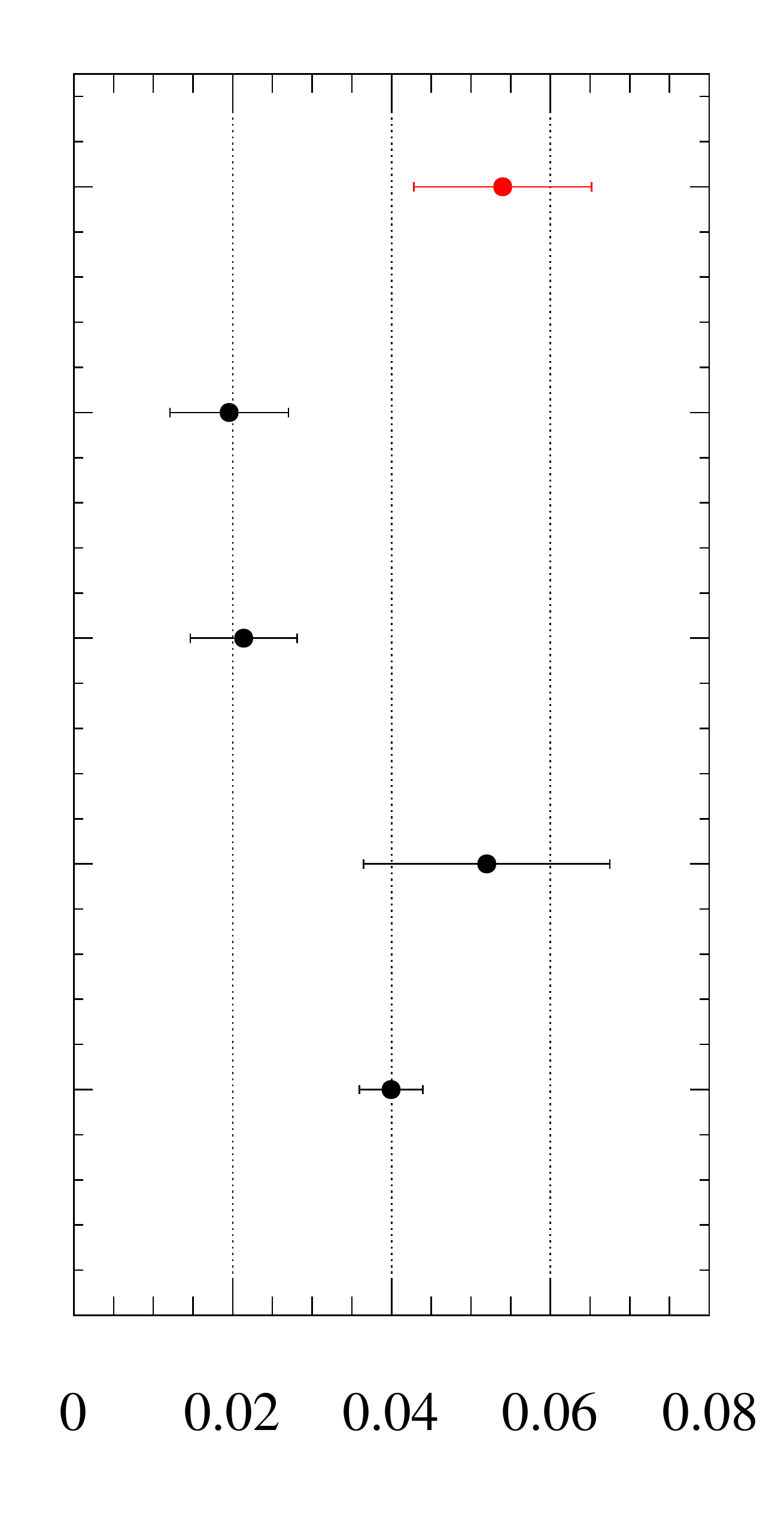}}
    \put(25,109){\large $\Lb\to\Ppsi\proton\Km$}
    \put(25, 91){\large $\Bz\to\Ppsi\Kstarz$}
    \put(25, 73){\large $\Bz\to\Ppsi\Kz$}
    \put(25, 55){\large $\Bu\to\Ppsi\Kz\pip$}
    \put(25, 37){\large $\Bu\to\Ppsi\pip$}

    \put( 85,5)  {\large$R_\psitwos$}
  \end{picture}
  \caption {\small Comparison of the branching fraction ratios
  measured in this analysis (red dot) with analogous results 
  obtained in neutral and charged B-meson decays~\cite{PDG2018} 
  to the \psitwos final-state. 
 The symbol \Ppsi in decay chains denotes
 a~$\psitwos$ or $\chiconex$~meson.}
  \label{fig:ratios_psi}
\end{figure}

\clearpage

\subsection*{Dipion mass spectra comparison}

\begin{figure}[hbt]
	\setlength{\unitlength}{1mm}
	\centering
	\begin{picture}(150,120)
	\put(0,0){\includegraphics*[width=150mm,height=120mm]{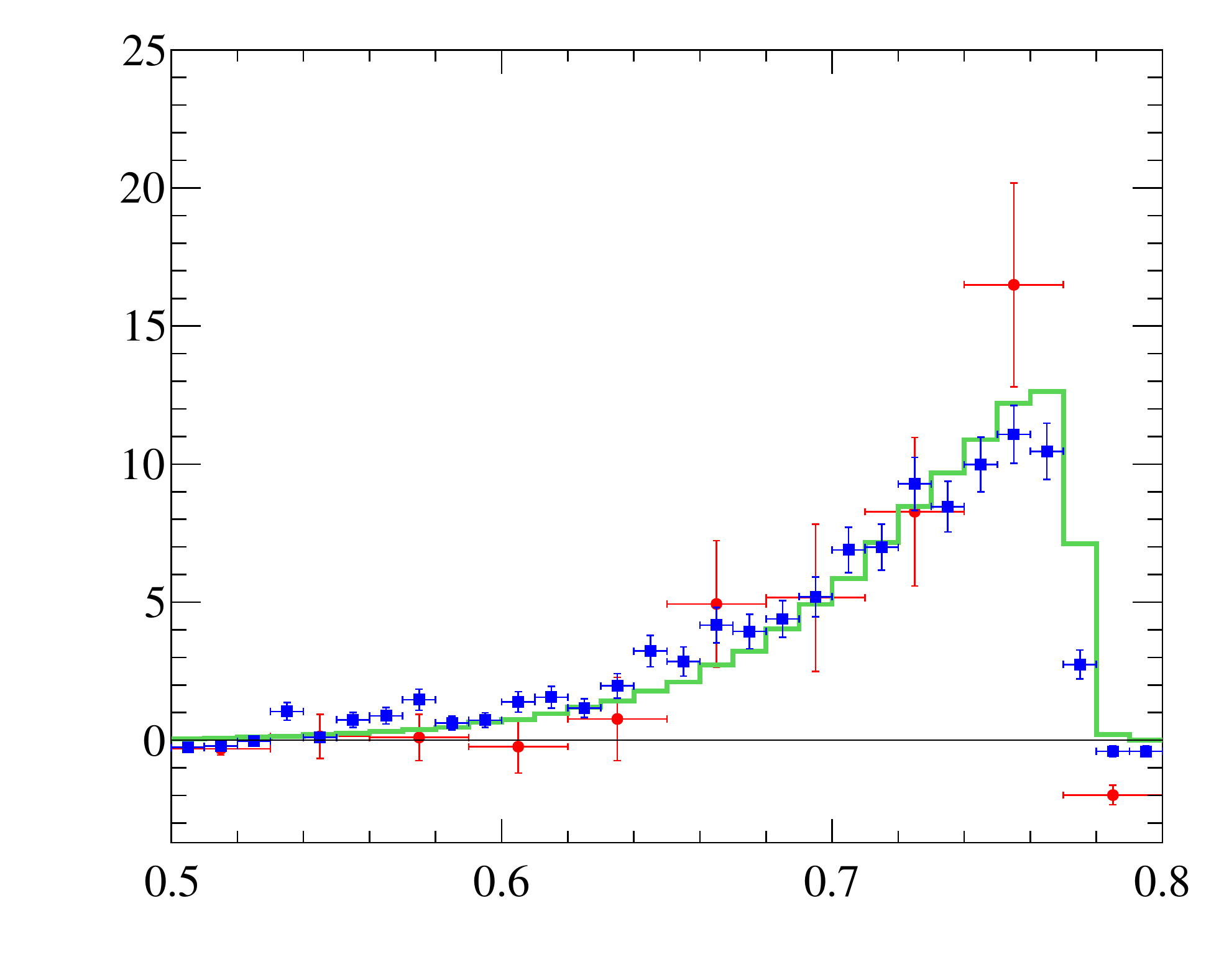}}
	\put(  1, 45){\Large\rotatebox{90}{Arbitrary units}}
	\put( 70,  2){\Large$m_{\pip\pim}$}
	\put(123,  2){\Large$\left[\!\gevcc\right]$}
	\put(120,103){\Large\lhcb}
	\put(28,105){\color[RGB]{89,212,84}     {\rule{6mm}{2.0pt}}} 

	\put(35,104){{$\XToRhoz $ simulation}}
	\put(35,96){{$\LbToXPK$ data}}
	\put(35,88){{\decay{\Bu}{\chiconex\Kp} data}}
	\put(29,97){\color{red}\line(1,0){4}} 
	\put(31,95){\color{red}\line(0,1){4}} 
	\put(30.5,99){\color{red}\line(1,0){1}} 
	\put(30.5,95){\color{red}\line(1,0){1}} 
	\put(29,96.5){\color{red}\line(0,1){1}} 
	\put(33,96.5){\color{red}\line(0,1){1}} 
	\put(31,97){\color{red}\circle*{1.5}}
	\put(29,89){\color{blue}\line(1,0){4}} 
	\put(31,87){\color{blue}\line(0,1){4}} 
	\put(30.5,91){\color{blue}\line(1,0){1}} 
	\put(30.5,87){\color{blue}\line(1,0){1}} 
	\put(29,88.5){\color{blue}\line(0,1){1}} 
	\put(33,88.5){\color{blue}\line(0,1){1}} 
	\put(30.2,88.3){$ \crule[blue]{1.5mm}{1.5mm}$}
	
	\end{picture}
	\caption {\small Background-subtracted mass distribution of the~${\pip\pim}$ combinations in (red)~\mbox{\LbToXPK} 
	and (blue)~\mbox{\decay{\Bu}{\chiconex\Kp}}~\cite{LHCb-PAPER-2015-015} decays. The distributions are normalized to equal area.
	}
	\label{fig:signalalt}
\end{figure}

Background-subtracted mass distribution of the~${\pip\pim}$ combinations is shown in Fig.~\ref{fig:signalalt}.

\clearpage
 \subsection*{Mass spectra for the~\boldmath{\chiconex\Km}~and
 \boldmath{\chiconex\proton} combinations from
 \boldmath{$\decay{\Lb}{\Pchi_{\cquark1}\mathrm{(3872)}\proton\Km}$}~decays}

\begin{figure}[ht]
	\setlength{\unitlength}{1mm}
	\centering
	\begin{picture}(150,60)
	
	\put( 0,0){\includegraphics*[width=75mm,height=60mm]{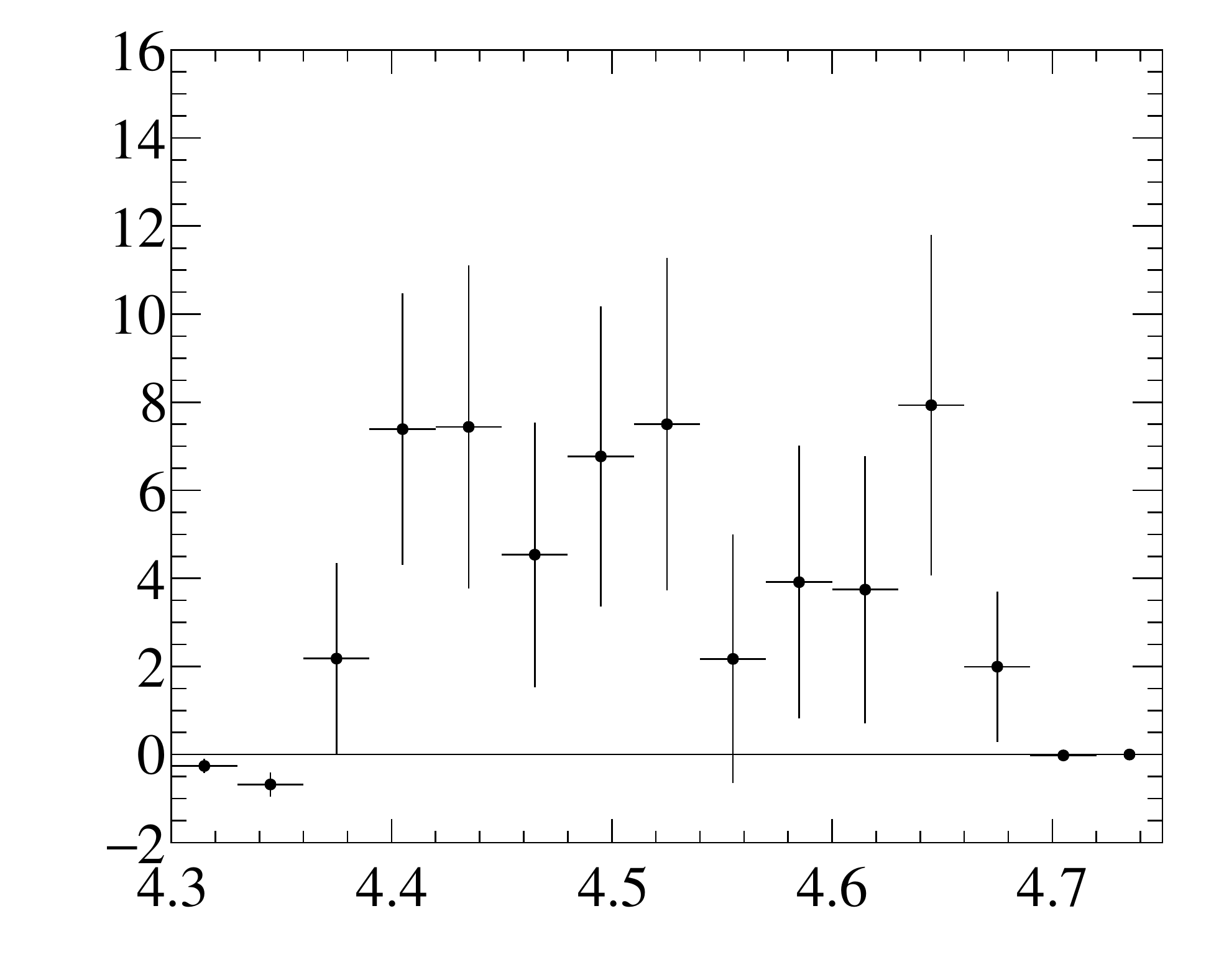}}
	\put( 80,0){\includegraphics*[width=75mm,height=60mm]{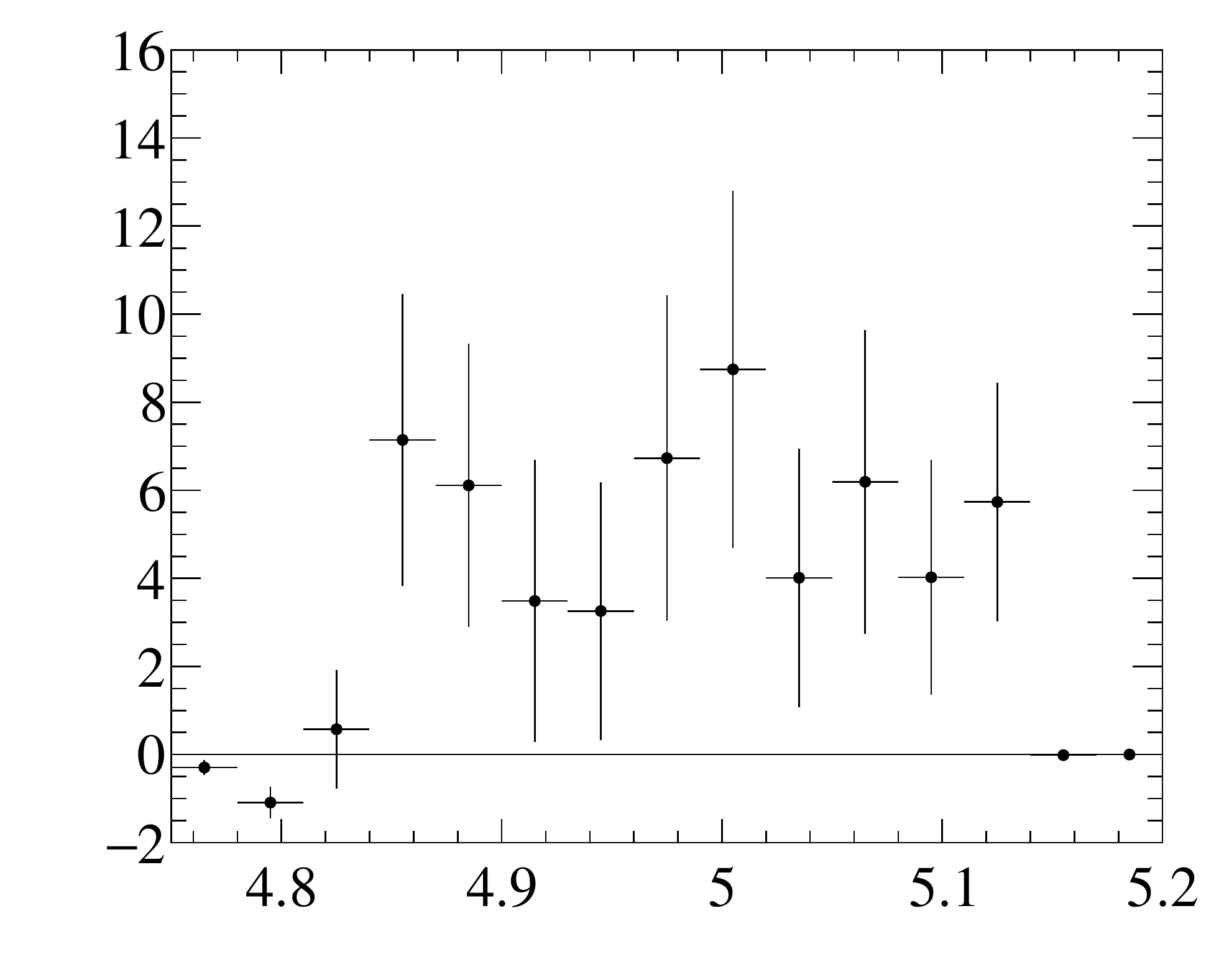}}

	\put(0,7){\begin{sideways}$\rm{N_{\LbToXPK}}/(30\mevcc)$\end{sideways}}
    \put(80,7){\begin{sideways}$\rm{N_{\LbToXPK}}/(30\mevcc)$\end{sideways}}

	\put(30,0){$m_{\chiconex\Km}$}
	\put(57,0){$\left[\!\gevcc\right]$}
	
	\put(112,0){$m_{\chiconex\proton}$}
	\put(137,0){$\left[\!\gevcc\right]$}
	
	\put( 55,50){\lhcb}
	\put( 135,50){\lhcb}

	\end{picture}
	\caption {\small Background-subtracted mass distributions for
        (left)~the~\chiconex\Km
        and (right)~\chiconex\proton~systems in \LbToXPK~decays.}
	\label{fig:last}
\end{figure}

Background-subtracted mass distributions for
the~\chiconex\Km and \chiconex\proton~systems are shown in Fig.~\ref{fig:last}.

\addcontentsline{toc}{section}{References}
\bibliographystyle{LHCb}
\bibliography{main,standard,LHCb-PAPER,LHCb-CONF,LHCb-DP,LHCb-TDR}